\documentclass[conference]{IEEEtran}
\IEEEoverridecommandlockouts

\usepackage{cite}
\usepackage{amsmath,amssymb,amsfonts}
\usepackage{algorithmic}
\usepackage{graphicx}
\usepackage{textcomp}
\usepackage{xcolor}

\usepackage[T1]{fontenc}
\usepackage{inconsolata}

\usepackage{relsize}

\usepackage{comment}
\usepackage{xspace}
\usepackage{multirow}
\usepackage{makecell}
\usepackage{array}
\usepackage{booktabs}
\usepackage{tabularx}
\usepackage{adjustbox}
\usepackage{pifont}
\usepackage{hhline} 
\usepackage{float}
\usepackage[table]{xcolor}
\usepackage[normalem]{ulem}
\usepackage{etoolbox}
\usepackage{ragged2e}
\usepackage{colortbl}
\usepackage{multicol}
\usepackage{listings}
\definecolor{darkgreen}{RGB}{0, 130, 0}
\usepackage{enumitem}
\usepackage{dashrule}
\usepackage{mdframed}
\usepackage{minted}
\usepackage{marginnote}
\usepackage[dvipsnames]{xcolor}
\usepackage{tcolorbox}
\tcbuselibrary{skins}
\usepackage{url}
\usepackage{hyperref}

\usepackage[scaled=1]{newtxtt} %
\usepackage{needspace} %

\newmdenv[
  linewidth=0.8pt,
  innertopmargin=2pt,
  innerbottommargin=2pt,
  innerleftmargin=2pt,
  innerrightmargin=4pt,
  skipabove=4pt,
  skipbelow=0pt,
]{finding}

\newcounter{rnote}

\usepackage[final]{microtype}
\hypersetup{hidelinks}

\newcounter{snote}

\newcolumntype{G}{>{\color{ForestGreen}}c}

\newcommand{\customsize}{\fontsize{7.8}{8.0}\selectfont}
\definecolor{rowgray}{gray}{0.93}
\definecolor{lightgray}{gray}{0.9}
\definecolor{mediumgray}{gray}{0.8}

\definecolor{lightblue}{RGB}{210,233,242}
\colorlet{ObsBlue}{RoyalBlue!50}

\newcounter{obscounter}
\newcommand{\obs}{\stepcounter{obscounter}\tcbox[on line, boxsep=0.0pt, left=0pt, right=0pt, top=0pt, bottom=0pt, colback=lightgray, colframe=lightgray]{\small\textbf{Obs. \theobscounter}}}

\newcounter{obscounterNEW}

\newcommand{\approach}{ITHICA\xspace}

\newcommand{\implication}[2]{\vspace{6pt}\par\noindent\textbf{\textit{#1}} \textit{#2}\vspace{6pt}}

\newcommand{\myparagraph}[1]{\par \textbf{\textit{#1.}}}

\newcommand{\cpucheck}{cpu-check}

\newcommand{\Company}{Google}

\definecolor{MyDeepGreen}{RGB}{60, 180, 75} 

\newcommand{\CC}{\texttt{CC}}
\newcommand{\CCithica}{\texttt{CC-ITHICA}}

\newcommand{\Native}{\texttt{Native}}
\newcommand{\ITHICA}{\texttt{ITHICA}}

\newcommand{\FB}{\texttt{FB}}

\newcommand{\FBithica}{\texttt{FB-ITHICA}}

\newcommand{\A}{\texttt{Arith}}
\newcommand{\M}{\texttt{Mem}}
\newcommand{\MD}{\texttt{MemDiv}}
\newcommand{\B}{\texttt{Br}}
\newcommand{\AM}{\texttt{Arith+Mem}}
\newcommand{\AMD}{\texttt{Arith+MemDiv}}
\newcommand{\AMDB}{\texttt{Arith+MemDiv+Br}}

\definecolor{Hgray}{RGB}{235,233,226}
\definecolor{Hgreen}{RGB}{14,100,14}
\definecolor{Hamber}{RGB}{180,90,0}
\definecolor{Hred}{RGB}{180,20,20}

\newcommand{\whitexmark}{\cellcolor{white} \raisebox{-0.3ex}{\textcolor{black}{\xmark}}}
\newcommand{\bluecmark}{\cellcolor{lightblue} \raisebox{-0.3ex}{\textcolor{black}{\cmark}}}

\newcolumntype{L}{>{\RaggedRight\arraybackslash}m{6.209cm}} %
\newcolumntype{S}{>{\RaggedRight\arraybackslash}m{4.73cm}} %
\newcolumntype{Y}{>{\centering\arraybackslash}X} %

\newcommand{\thickhline}{\Xhline{0.8pt}} %

\newcommand{\cmark}{\ding{51}}
\newcommand{\xmark}{\ding{55}}

\newcolumntype{C}[1]{>{\centering\arraybackslash}p{#1}}

\def\BibTeX{{\rm B\kern-.05em{\sc i\kern-.025em b}\kern-.08em
    T\kern-.1667em\lower.7ex\hbox{E}\kern-.125emX}}

\begin{document}
\flushbottom

\title{ITHICA: Intra-Thread Instruction Checking
Approach for Defect-Induced Silent \\ Data Corruptions 
}

\author{\IEEEauthorblockN{Ioanna Vavelidou}
\IEEEauthorblockA{\textit{Stanford University} \\
vavel@stanford.edu}
\and
\IEEEauthorblockN{Subho Sankar Banerjee}
\IEEEauthorblockA{\textit{Google LLC} \\
subhob@google.com}
\and
\IEEEauthorblockN{Eric Xue Liu}
\IEEEauthorblockA{\textit{Google LLC} \\
ericxliu@google.com}
\and[\hfill\mbox{}\par\mbox{}\hfill]
\IEEEauthorblockN{Mike Fuller}
\IEEEauthorblockA{\textit{Google LLC} \\
mikefuller@google.com}
\and
\IEEEauthorblockN{Subhasish Mitra}
\IEEEauthorblockA{\textit{Stanford University} \\
subh@stanford.edu}
\and
\IEEEauthorblockN{Caroline Trippel}
\IEEEauthorblockA{\textit{Stanford University} \\
trippel@stanford.edu}
}

\maketitle

\begin{abstract}
Hyperscalers are reporting  silent data corruptions~(SDCs), presumed to be caused by silicon manufacturing defects, %
as a threat to datacenter reliability. To support datacenter testing efforts to detect defective CPU servers, this paper 
presents \textit{ITHICA}, an approach and tool for automatically generating functional tests for defect-induced errors from arbitrary programs by inserting intra-thread, instruction-level error checks, primarily leveraging instruction duplication and output comparison.
Our key insight %
is that the most pernicious defects (those most likely to escape manufacturing testing) cause \emph{inconsistent} errors: %
two executions of the same instruction  given the same inputs within the same thread can %
produce \emph{different} architectural outputs depending on the execution context in which they run. 
By exploiting this insight, ITHICA %
uniquely
enables arbitrary programs to serve as tests and localizes %
affected instructions %
concurrently with error detection. %
We use ITHICA to transform industrial hyperscaler tests (our baseline), datacenter programs, and common libraries into functional tests, and evaluate them on over 3,000
CPU servers.  ITHICA checks detect 39\% more defective servers than baseline industrial checks and yield novel findings on defect behavior that challenge conclusions drawn by 
prior hyperscaler~fleet~studies.
\end{abstract}


\section{Introduction}
\label{sec:intro}

Hyperscalers are reporting \textit{silent data corruptions} (SDCs), presumed to be caused by silicon \textit{manufacturing defects}, resulting in permanent (hard) faults, as a critical threat to datacenter reliability~\cite{Dixit:2021:sdc-at-scale, dixit:sdc-detecting, Hochschild:2021:cores-dont-count, Wang:alibaba:sdc, alibaba-new,10xmitra,IEEE-Micro-growing-concern}.
Such SDCs occur when a \textit{hardware error}~(\S\ref{subsec:symptoms}) causes a system to output an incorrect result without any indication that the error occurred~\cite{myth-or-reality}. 
\looseness=-1
Hyperscalers are finding that defect-induced SDCs  
impact a substantial fraction of deployed hardware, roughly one silicon device per thousand~\cite{dixit:sdc-detecting,Dixit:2021:sdc-at-scale,Hochschild:2021:cores-dont-count,Wang:alibaba:sdc,10xmitra}.

To detect defective hardware, hyperscalers have begun subjecting their server fleets to frequent functional testing. This testing has enabled hyperscalers to publish several fleet-wide studies characterizing the SDC problem~\cite{Dixit:2021:sdc-at-scale,dixit:sdc-detecting,Hochschild:2021:cores-dont-count,Wang:alibaba:sdc,10xmitra,alibaba-new,pindrop}, but these efforts
rely on proprietary, hyperscaler- or vendor-specific tests. While some tests %
have been open-sourced by hyperscalers and %
vendors~\cite{cpu-check,intel:opendcdiag,sevi-meta}, %
these public suites are generally smaller in scope, and vendor tests especially tend to be structurally distinct from proprietary variants, %
making them difficult to build on and improve. 

Two bodies of work attempt to overcome the limitations of public functional test suites through automated test generation. The first 
generates tests that expose defects as captured by a specific fault model on a specific hardware implementation~\cite{vega}, %
but is limited by the inaccuracy of current fault models for defects~\cite{McCluskey:stuck-fault,Wei:pepr,10xmitra,intel-screening-manuf-defects}
and the scalability challenges of fault-model-driven test generation~\cite{mahesh:rtl-stuck-at-bmc, Kundu:blast}.
The second employs fuzzing to generate tests that thoroughly exercise a specific hardware implementation~\cite{google:silifuzz,harpocrates,harpocrates_plusplus}. 
Both approaches generally require below-ISA hardware specifications, which are hardware-vendor-proprietary and unavailable to end-users including hyperscalers, making them inapplicable for independent, in-datacenter fleet evaluation~\cite{google:silifuzz}. 
The one exception is SiliFuzz~\cite{google:silifuzz}, which %
requires ISA-level specifications only,  but generates short 
tests that have been deemed less effective than proprietary tests~\cite{10xmitra}, a finding we corroborate in \S\ref{fuzzing}.

\myparagraph{The case for  programs as tests}
Exercising hardware diversely is essential for triggering defect-induced errors, but without below-ISA specifications end-users cannot control or measure hardware coverage~\cite{google:silifuzz}. Instead, this paper  proposes heuristically approximating this goal by \emph{deriving functional tests from arbitrary programs}.
This strategy is well motivated: real datacenter programs expose defects that synthetic tests miss~\cite{10xmitra,hardware-sentinel-asplos25}, by exercising complex microarchitectural features needed to trigger defects in ways that are difficult for synthetic tests to imitate~\cite{pindrop}.

\myparagraph{The barrier imposed by existing checks}
The key to transforming arbitrary programs into functional tests is instrumenting them with checks for defect-induced errors. Looking to recent work 
that describes or publishes functional test content~\cite{intel:opendcdiag,cpu-check,google:silifuzz,harpocrates,harpocrates_plusplus,Wang:alibaba:sdc,alibaba-new,pindrop,sevi-meta,Dixit:2021:sdc-at-scale,dixit:sdc-detecting},  %
three methods are used: comparing 
\looseness=-1
(i) inputs and outputs of invertible computations~(e.g.,~encryption followed by decryption)~\cite{cpu-check,intel:opendcdiag};  (ii) outputs of the same computation run on different threads/cores~\cite{cpu-check,intel:opendcdiag,google:silifuzz,pindrop}; %
and (iii) the output of a computation to a 
golden value computed on presumed-healthy hardware~\cite{intel:opendcdiag,google:silifuzz,sevi-meta,pindrop,harpocrates,harpocrates_plusplus,alibaba-new,Wang:alibaba:sdc,Dixit:2021:sdc-at-scale,dixit:sdc-detecting}. 
Notably, none of these methods compares the architectural outputs of the same instruction given the same architectural inputs within the same~thread.

Unfortunately, these error checking methods  severely limit 
which programs can serve as tests. Few are 
composed of invertible computations, and most 
lack well-defined final outputs
for golden-value or cross-thread/core checks. Even when such outputs exist, these checks require deterministic execution, making them impractical for programs involving multi-threading, network I/O, or %
other sources of non-determinism.
These challenges are amplified if finer-grained (i.e., more frequent) checking is desired in order to %
reduce logical error masking 
or localize errors to instructions. %
For this reason, prior work  
performs instruction localization  using 
short %
tests only~\cite{sevi-meta} %
or by heuristically attributing errors to instructions that frequently occur 
across failing~tests~\cite{alibaba-new, Wang:alibaba:sdc}.

\myparagraph{This Paper: Enabling programs as tests with ITHICA}
Instead, we propose 
\approach{} (Intra-THread Instruction Checking Approach), 
 an automatic approach and tool that generates functional tests %
from programs by instrumenting them with \textit{intra-thread}, \textit{instruction-level} checks,
primarily leveraging \textit{instruction-duplication and output comparison}~\cite{EDDI, CFCSS, Swift, Hong:2010:qed,lin2013overcoming,Singh:2017:eqed,Lin2014Effective}.
These checks eliminate the need for final program outputs and significantly reduce determinism requirements, enabling many more programs to serve as tests. Their instruction granularity makes them less susceptible to error masking and enables %
identification of failing instructions upon
detection.

\approach{}'s %
checking approach is enabled by our \textbf{key insight}: %
the most \textit{pernicious} defects---those most likely to escape manufacturing testing into production---cause \textit{inconsistent} errors: two instances of the same instruction given the same architectural inputs within the same thread can \textit{sometimes} %
produce \textit{different} architectural outputs.
Intuitively, such inconsistencies result from extensive microarchitectural and electrical state in modern processors, which creates a highly diverse execution context for each dynamic instruction instance, thereby modulating how defects affect each instance architecturally. 
In contrast to \textit{consistent} errors---two instances of  the same instruction given the same architectural inputs within the same thread \textit{always} produce the \textit{same} wrong architectural output---that can be exposed from \textit{every execution context}, %
inconsistent errors are sensitive to specific contexts drawn from an overwhelming space of possibilities~(\S\ref{sec:qed-theory}).
\looseness=-1
Notably, inconsistent errors can arise even when a  defect affects both instruction instances; %
for example, on one server we evaluate~(\S\ref{Results4}), two identical instructions that strongly appear to interact with the same faulty hardware component produce \textit{different wrong outputs} (a type of inconsistent error), enabling ITHICA to detect the errors. %

Our \textbf{first contribution} is \approach{} itself. To our knowledge, it is the first work to 
validate that defects cause inconsistent errors on real defective hardware and to exploit this phenomenon for defect detection. \approach{} primarily repurposes intra-thread, instruction-level checking techniques from prior work on soft error detection~\cite{EDDI, CFCSS, Swift} and post-silicon validation~\cite{Hong:2010:qed,lin2013overcoming,Lin2014Effective,Singh:2017:eqed}. Additionally, it introduces a novel technique  
that proactively diversifies the execution context of memory instructions, by encouraging them to interact with different levels of the
memory hierarchy. %
Implemented as LLVM compiler passes, \approach{} can be applied to arbitrary programs across multiple architectures.
By relying neither on proprietary specifications nor fault models, it bridges the research gap between state-of-the-art functional tests deployed in datacenters and what is publicly available~to~researchers.

Our \textbf{second contribution} is a large-scale testing campaign that demonstrates \approach{}'s effectiveness. 
From \Company{}'s industrial fleet of multiple millions of CPU servers, over 3,000 suspect- and confirmed-defective servers---spanning at least 10 microarchitectures---were identified and subjected to multiple \approach{} test types:
derived from representative open-source hyperscaler test programs~\cite{cpu-check} (our  baseline), fleet-representative workloads~\cite{fleetbench}, and common libraries~\cite{zlib,openssl,abseil,libcllvm,libcxxllvm}. %
ITHICA detects \textit{100 defective servers}, which
we analyze in detail---more than 
3$\times$~\cite{Wang:alibaba:sdc} and 5$\times$~\cite{sevi-meta} the number
of the two most detailed prior studies.
By exploiting inconsistent errors,
\approach{} checks detect 39\% more defective servers than \textit{native} final output checks within the same binaries, and 69\% more than SiliFuzz~\cite{google:silifuzz}, the only other functional test generation approach operating at or above the ISA level. ITHICA is the first technique to be directly compared against open-source hyperscaler tests on real defective hardware.

Our \textbf{third contribution} is a set of novel findings on defect behavior uniquely enabled by \approach{}'s instruction-level error checking %
within a variety of programs and the high number of servers these tests detect. Many of these findings challenge conclusions drawn by recent hyperscaler studies:  
\begin{enumerate}
\item \approach{}'s unique ability to perform instruction localization 
within arbitrarily long tests enables our discovery that 
the \textit{sequence-driven execution context} in which a vulnerable instruction executes is the primary predictor of error manifestation~(\S\ref{Results3}). In contrast, \textit{instruction usage stress} (i.e., the dynamic frequency of a failing opcode in a test)~\cite{Wang:alibaba:sdc,alibaba-new} and \textit{culprit inputs} are insufficient  predictors~(\S\ref{Results4}). 
\item \approach{} enables empirically demonstrating the extreme difficulty of reproducing errors with short  tests: among~the defective servers  analyzed, \textit{only one reproduces errors with single-instruction tests}; the rest require sequences ranging from modestly longer to full programs~(\S\ref{Results4}). Prior~work characterizing error trends  based primarily  on those reproducible with short 
tests~\cite{sevi-meta}  may bias error characterization
towards the few cases for which short sequences are effective.
\item \approach{}'s ability to test a wide variety of instructions in diverse execution contexts enables our finding that the same defect often causes errors \emph{across multiple instruction types at markedly different rates}---up to six orders of magnitude~(\S\ref{Results5}). 
In one such case, the detecting hyperscaler baseline test is one targeting vector instructions, yet \approach{} also reveals failing non-vector instructions within the same test program. This finding cautions against broad claims of \textit{hardware localization} (i.e., attributing defects to hardware components) with functional tests in general, especially those that target few instruction types~\cite{sevi-meta,alibaba-new,Wang:alibaba:sdc}.
\end{enumerate}

\looseness=-1
Our findings suggest effective SDC testing requires instruction-level checking within long, diverse programs. The ideal---exercising every instruction in every possible execution context---remains an open challenge. ITHICA takes a meaningful step towards it: by %
leveraging the inconsistent nature of defect-induced errors,  
it opens up a vast space of previously unusable programs as tests, each exposing instructions to execution contexts that prior checking methods cannot leverage.

\section{Background} %

\subsection{Hardware Faults}
\label{sources}
Hardware faults are physical flaws or malfunctions. They can be permanent or transient.

\myparagraph{Permanent Faults} 
\label{subsubsec:permanent}
\textit{Permanent (hard) faults} include manufacturing defects,
aging-related wear-out, and design bugs.

\textit{Manufacturing defects} %
are flaws introduced during silicon chip manufacturing,
and typically impact different physical regions of each affected chip in different ways.
A special category of defects is \textit{early-life failures} (ELFs)%
, which are
characteristic of weak chips that pass pre-deployment testing~\cite{ELF-kim13,10xmitra}, %
but induce 
erroneous %
behaviors within the first few weeks to months of deployment~\cite{gate-oxide-pred}.

Distinct from defects (\textit{extrinsic} flaws)~\cite{ELF-kim13}, \textit{aging-related faults} are caused by \textit{intrinsic} failure mechanisms (e.g., 
Bias Temperature Instability (BTI)~\cite{BTI} or Hot Carrier Injection~(HCI)~\cite{HCI}), which result from chip wear-out over time. 
Chip designs incorporate voltage and speed margins %
to 
prevent errors from aging faults~\cite{aging-mitra}. %

Finally, electrical and logic \textit{design bugs} are %
flaws in a \textit{design};
thus, they affect most or all chips in a manufacturing batch.
Electrical bugs manifest under specific operating conditions, namely voltage, frequency, and/or temperature~\cite{Patra07}.
Logic bugs are flaws in a chip's logical implementation.

\myparagraph{Transient Faults}
\label{subsubsec:transient}
\textit{Transient faults} 
 are malfunctions that
occur when energetic particles, such as neutrons from cosmic rays~\cite{ziegler1979effect}~or alpha particles from packaging materials~\cite{PhysMech-SoftErrors}, %
generate electron-hole pairs in semiconductor devices. The resulting charge can accumulate at a transistor’s %
diffusion 
nodes and %
flip~the~state of a logic device, a
phenomenon referred to as \textit{transient (i.e., soft)}~error.
Transient faults are %
\textit{not} the result of permanent hardware flaws~\cite{Mukherjee:soft-error-problem-arch}.

\subsection{Symptoms of Hardware Faults} 
\label{subsec:symptoms}
A hardware fault can induce a \textit{hardware error}, i.e., an incorrect bit stored in a flip-flop.
Whether a permanent fault %
induces a hardware error may be: \textit{sequence-dependent}, i.e., dependent on the order of circuit stimuli; or
\textit{timing-dependent}, i.e., a subset of sequence-dependent that depends on the speed of  stimuli~\cite{diagnosis-sequence-dependent}. Timing-dependent faults are dependent on a chip's electrical state (i.e., voltages and currents at transistor terminals) and thus influenced by external factors like power supply noise, frequency  or temperature variations~\cite{Chang1998patterndependent,delay-avf}.%

\looseness=-1
A hardware error
may or may not manifest as an \textit{architectural error},  
i.e., an incorrect architectural state~(\S\ref{sec:qed-theory}).
If it does not,
the hardware fault is said to be
\textit{architecturally masked}~\cite{fault-tolerant-comparch-book} or \textit{benign}~\cite{arch-design-for-soft-errors,AVGI}.
Architectural errors can present with~various symptoms, including
system crashes, system hangs,
or
silent data corruption (SDC)~\cite{10xmitra,hardware-sentinel-asplos25}.
An SDC occurs when an architectural error causes the system to output an incorrect result without any indication that the error occurred~\cite{myth-or-reality}.
Some architectural errors may not affect the system output at all
due to \textit{logical masking}~\cite{arch-design-for-soft-errors,avf,AVGI,fault-tolerant-comparch-book},
e.g., if the error is multiplied by zero, or a program’s output %
does not depend on it.

\subsection{Testing Silicon Chips for Defects}
The SDC phenomenon recently reported by hyperscalers is generally attributed to %
defects%
~\cite{Dixit:2021:sdc-at-scale, dixit:sdc-detecting, Hochschild:2021:cores-dont-count, Wang:alibaba:sdc, alibaba-new}. Standard and emerging approaches for detecting defective chips are \textit{manufacturing} and \textit{in-datacenter} testing, %
respectively.
\label{sec:background:testing}

\label{sec:background:manufacturertest}
\myparagraph{Manufacturing Testing}
Manufacturing testing follows chip fabrication and 
subjects each chip to scan-based \textit{structural tests}~\cite{scan88} and (non-scan) \textit{functional tests}~\cite{Shen1998NativeMF}. %
Scan-based %
testing uses \textit{design-for-testability} (DFT) features to apply automatically generated test stimuli to logic circuits %
and inspect their outputs, leveraging fault models and test metrics for systematically generating scan tests~\cite{Wadsack:stuck-open, Waicukauski:transition-fault, Smith:path-delay-fault, Heragu:segment-delay-fault, Storey:bridging-fault,Ma:n-detect, Hapke:cell-aware-fault, McCluskey:gate-exhaustive-fault}. 
Functional testing complements scan testing,
but lacks scalably computable, high-quality coverage metrics~\cite{Kundu:blast,Wei:pepr}. %
Manufacturing testing is highly limited in duration
due to cost, making exhaustive testing approaches  infeasible~\cite{Wei:pepr}. %

 \myparagraph{In-Datacenter Testing} 
\label{sec:background:datacentertest}
\looseness=-1
Historically, in-datacenter testing concluded with %
software burn-in testing~\cite{dixit:sdc-detecting} (different from hardware burn-in~\cite{character-inf-mort}),
assuming hardware faults would be caught by built-in error detection mechanisms.
Upon finding that defect-induced SDCs impact a substantial fraction of  servers, datacenter
 operators have begun subjecting~them~to~periodic functional testing~(\S\ref{sec:intro}) both in- and out-of-production~\cite{Wang:alibaba:sdc,google:silifuzz,dixit:sdc-detecting}.

\section{Motivating ITHICA: The Inconsistent Manifestation of Pernicious Permanent Faults}
\label{key-insight}

\looseness=-1
\textbf{Our key insight} in designing \approach{}
is two-fold: %
(1) defects~can cause \textit{inconsistent} architectural errors, i.e., errors that 
cause some instruction, executed from two different \textit{execution contexts} on the same hardware, with identical architectural inputs, to produce \textit{different} architectural outputs; and (2)  such defects are the most pernicious. %
An execution context denotes the entire microarchitectural (including architectural) and electrical state of a chip.
We first %
derive this insight 
from a recent theoretical result (\S\ref{sec:qed-theory}).
We then build intuition for how permanent faults can cause inconsistent errors
with a pedagogical RTL-level fault injection example in simulation~(\S\ref{sec:rtl-fault-injection}).

\subsection{Architectural Errors: Formal Classification }
\label{sec:qed-theory}

Part one of our insight follows from recent work on formal pre-silicon verification %
(to detect logic bugs, a type of permanent fault,~\S\ref{subsubsec:permanent}), which proves that a hardware error can manifest as one of three types of architectural 
errors~\cite{theoretical-framework-sqed, GQED}:\footnote{Inconsistent, consistent, and unresponsive errors map to \textit{functional consistency}, \textit{single-action correctness}, and \textit{response bound} bugs in prior work~\cite{GQED}.} (1)~an \textbf{inconsistent error}, defined above; (2)~a \textbf{consistent error}, which causes some 
 instruction, executed from every execution context on the same hardware device, to always produce \textit{the same} wrong output for specific architectural inputs; or (3)~an~\textbf{unresponsive error}, which causes some instruction, executed from some context, to fail to produce an output within the expected time frame.

Notably, this result holds regardless of how the hardware error arises, e.g., even if it is caused by a fault that is both timing- and sequence-\textit{independent}.
Now, consider such a fault alongside one that is timing- or sequence-\textit{dependent},
where both faults are capable of inducing the same hardware error~(\S\ref{subsec:symptoms}). That is, the first fault induces the hardware error \textit{persistently}, while the second does so only under specific conditions~(i.e., \textit{intermittently}).
If the error manifests as inconsistent or unresponsive in the first case, it will do so in the second.
If the error manifests consistently in the first case, it will
newly manifest inconsistently in the second.%

Part two of our insight is that faults that induce inconsistent errors are more pernicious (harder to detect), and thus more  likely to escape manufacturing testing, than those that induce consistent errors. This is because consistent errors can be exposed by executing instructions with diverse inputs from \textit{any} execution context, while inconsistent errors require executing instructions in \textit{many} distinct execution contexts, drawn from an \textit{overwhelming} space of possibilities.

\begin{figure*}[t]
    \centering
\includegraphics[width=1\linewidth]{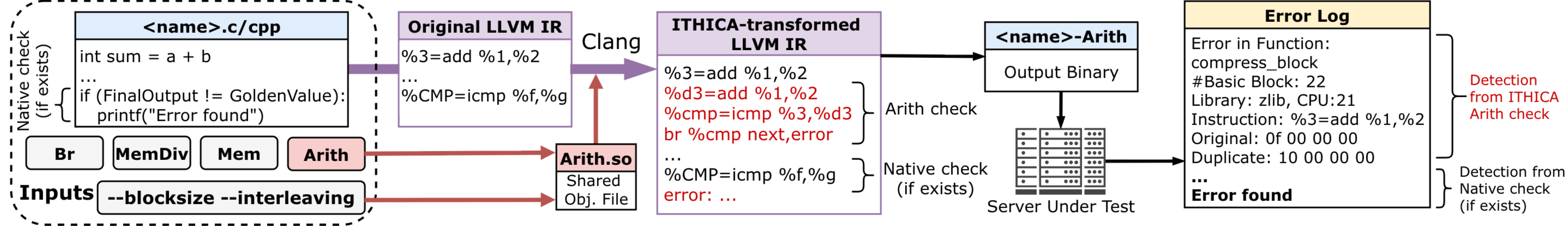}
    \caption{Given an input program (\texttt{<name>.c/cpp}), 
\approach{} applies one or more transformations---implemented in this paper for LLVM IR---configured with some  \texttt{blocksize} and  \texttt{interleaving}, and outputs a functional test \texttt{<name>-ITHICA} with intra-thread instruction  checks.
}
\label{fig:flow}
\end{figure*}

\approach{} is 
explicitly 
designed to detect inconsistent errors and can implicitly detect unresponsive errors, though not consistent ones.
This tradeoff pays off empirically: \approach{} detects more defective hardware---that has already undergone manufacturing testing---than state-of-the-art tests that do not exploit this insight~(\S\ref{results}).

\subsection{Pedagogical RTL Fault Injection Example}
\label{sec:rtl-fault-injection}
To build intuition for the theory in \S\ref{sec:qed-theory}, we present %
an RTL-level fault injection example %
featuring a timing- and sequence-independent fault---the case where consistent errors are theoretically more prevalent, and thus  where ITHICA is \textit{least} likely to detect an inconsistent error.

\looseness=-1
Our fault injection targets the open-source RISC-V CVA6 processor~\cite{cva6}, a 64-bit, 6-stage, single-, in-order-issue core with speculation. %
We inject a  single (i.e., one-bit) stuck-at fault~\cite{McCluskey:stuck-fault}---a common
 fault model used in the literature to study 
permanent 
faults---into CVA6's SystemVerilog RTL.
Specifically, %
we force \texttt{rs1\_valid\_o} permanently high, causing a persistent hardware error.
This signal informs the next instruction to issue that its first operand (\texttt{rs1}) is valid.
We then run a simple \approach{} test on the ``faulty'' design in %
Verilator simulation~\cite{verilator}.
The test is  %
an assembly program that features
two multiplication instructions (\texttt{mul}) using the same architectural inputs,
preceded by initialization instructions,  
as shown next: %

\definecolor{codecomment}{HTML}{3D7B7B}
\needspace{10\baselineskip}
\begin{lstlisting}[language=C, escapeinside=||,
  basicstyle=\linespread{1.06}\ttfamily\customsize,
  commentstyle=\color{codecomment}\itshape,
  numbers=left, numberstyle=\color{gray}\ttfamily\customsize,
  numbersep=8pt, xleftmargin=9.96pt, belowskip=2pt]
lw a5, |\textcolor[gray]{0.4}{\%}|[bval] // Sets value for a5
lw a4, |\textcolor[gray]{0.4}{\%}|[aval] // Sets value for a4
addi t0, a5, |\textcolor[gray]{0.4}{0}| // Dummy instruction
|\textcolor{blue}{mul}| |\textcolor[gray]{0.4}{\%}|[s1], a4, a5 // Original mul instruction
|\textcolor{blue}{mul}| |\textcolor[gray]{0.4}{\%}|[s2], a4, a5 // Duplicate (validation) mul instruction
\end{lstlisting}

We compare the original and duplicate \texttt{mul}s' outputs once the simulation has completed.  
By forcing \texttt{rs1\_valid\_o} permanently high,  both \texttt{mul}s are forced to perform forwarding of their~\texttt{rs1} operand value, without explicitly waiting for their producer to complete.
Getting the value from the register file is not permitted, due to the presence of the in-flight \texttt{lw} producer~(line 2) that writes to the same operand. As a result, the first \texttt{mul} 
forwards a reset value (zero) for \texttt{rs1}, which was written in the scoreboard when the \texttt{lw} instruction issued. 
The second \texttt{mul} %
forwards the correct value from the \texttt{lw} (line~2), which has completed. %
The  same effect is achieved if the \textit{dummy} instruction %
is placed between the two \texttt{mul} instructions instead.

Notably, the microarchitectural simplicity of CVA6 (and the abstractness of RTL) %
makes such a persistent hardware error  \textbf{more challenging} for ITHICA to detect: 
there is \textit{less} opportunity to observe a resulting inconsistent architectural error. %
Any increase in microarchitectural complexity makes inconsistent errors easier to observe, e.g., a processor with multiple execution units of the same type may dispatch the original and duplicate instructions to different units. %
However, execution unit redundancy %
is not a requirement: our real hardware results %
feature a case of an instruction %
exhibiting inconsistent errors, detected by ITHICA, despite very likely %
executing on the same faulty execution unit (\S\ref{Results4}).

\subsection{Execution Context Diversity}
\label{execution-context-diversity}

ITHICA exploits our key insight~(\S\ref{key-insight}) for defect detection by leveraging \textbf{natural diversity}, where the execution context varies passively between the original instruction and its validation (e.g., duplicate in~\S\ref{sec:rtl-fault-injection}) instruction~(\S\ref{sec:configuring}), and \textbf{proactive diversity} via the novel \MD{} transformation~(\S\ref{sec:ithica-passes}), which deliberately perturbs microarchitectural state,  and thus execution context, between  original and validation instructions. ITHICA also varies input programs~(\S\ref{sub:software-setup}) and transformations that turn the input programs into tests~(\S\ref{sec:ithica-passes}) to more significantly vary the execution context of both original and validation instructions, thus increasing the likelihood of surfacing defect-induced errors~(\S\ref{sec:qed-theory}).

\section{\approach{}: Intra-THread Instruction Checking Approach for Defect Detection}
\label{sec:ITHICA}

\looseness=-1
\approach{}\footnote{{%
\url{https://github.com/ioannavav/ithica}
}}  
takes as input a program and configuration options~(\S\ref{sec:configuring}) and outputs a functional test %
by applying one or more of its transformations~(\S\ref{sec:ithica-passes}) to the program~(Fig.~\ref{fig:flow}).
This section describes test generation with ITHICA and provides details on its %
design and implementation~(\S\ref{sec:llvm-implementation}).

\subsection{\approach{} Program Transformations}
\label{sec:ithica-passes}
Each ITHICA transformation (Table~\ref{tab:transformations}) inserts additional program instructions 
to check for errors in the outputs of specific
\textit{original} program instructions, including: (i) \textit{validation} instructions, which perform redundant computations; (ii) %
\textit{check} instructions, which detect inconsistent errors between original/validation instructions; (iii) \textit{error-reporting branches}, which redirect control flow to an error-reporting basic block when an inconsistent error %
is detected; 
and (iv) novel \textit{diversity} instructions, which create proactive diversity in microarchitectural state between the execution of original and validation instructions. 
The first three transformations below are adapted from prior work on
post-silicon validation for design bugs~\cite{Hong:2010:qed,lin2013overcoming,Lin2014Effective,Singh:2017:eqed} %
and soft error detection~\cite{EDDI, CFCSS, Swift}.   %

\label{sec:arith}
\textbf{\A{}} 
instruments arithmetic (i.e., computational) instructions. %
Validation instructions %
duplicate the original ones 
and their results are checked for mismatches against them. %
Duplicates preserve original data dependencies, allowing error propagation and detection %
even when comparisons are inserted less frequently (\S\ref{blocksize}).

\label{sec:mem}
\looseness=-1
\textbf{\M{}} instruments load and store instructions.
Loads are duplicated  and their outputs are compared against the originals.
For stores, validation instructions are loads from the store's address, and their outputs are compared against the original store's value operand.  %

\label{sec:branch}

\textbf{\B{}} instruments conditional branches to verify that they resolve to the correct target  %
by recording the \textit{intended} target address before the branch resolves, %
and validating it against the \textit{actual} %
address taken. %
\B{} %
is more lightweight than prior control-flow checking work~\cite{CFCSS, Swift, nZDC, flowery}: %
it exclusively targets conditional branches, %
and focuses on wrong-target misdirections between two valid targets, %
which are more likely to manifest silently. 
Branching to a rogue address is more likely to result in a non-silent architectural error, like a crash (\S\ref{subsec:symptoms}), which would be detected without %
 functional testing~\cite{chatzopoulos2025gates}.

A limitation of \M{}, shared by ITHICA and prior work~\mbox{\cite{Lin2014Effective,lin2013overcoming}},
is that validation loads typically retrieve data: via
an L1 cache hit (for checking loads), possibly missing errors in load outputs during L1 misses; or a core-local store buffer~(for checking 
stores), possibly missing errors in store outputs at lower memory hierarchy levels.
To address this, %
ITHICA introduces the novel \MD{} transformation, representing the first use of \textit{proactive diversity} (\S\ref{execution-context-diversity}) to check for %
errors affecting memory instructions. 

\textbf{\MD{}} extends \M{} by %
introducing diversity instructions, namely
\textit{mfence} (\textit{memory fence}) and \textit{clflush} (\textit{cache line flush})~(though others are also possible).
These %
instructions flush store buffers and caches by virtual address, respectively, encouraging original and validation memory instructions to interact with and exercise defects within different levels of the memory hierarchy. Specifically, %
\textit{mfence} encourages retrieval via an L1 cache hit. The subsequent \textit{clflush} forces data retrieval directly from main memory. 
Note that parity/ECC might not cover all data
movement of a complex memory system, nor be able to detect all errors~\cite{revisiting-memory-errors}.

\subsection{\approach{} Configuration Options}
\label{sec:configuring}
\approach{} supports configuring %
the location of validation instructions and the insertion frequency of comparison instructions per basic block. Configuration options are available for all transformations except \B{}, which targets at most one instruction per basic block.

\myparagraph{Configuring Interleaving} 
\label{conf-interleaving}
\approach{}'s \textit{interleaving} %
controls the insertion points of validation/diversity instructions relative to originals.
An interleaving of $n$ indicates that $n$ original instructions will appear consecutively in program order before their corresponding validation/diversity instructions appear %
in the same order.

\myparagraph{Configuring Block Size}
\label{blocksize}
\approach{}'s \textit{block size} %
controls the insertion frequency of comparison instructions.
A block size of $n$ indicates that comparisons will be inserted for every \mbox{$n$-th} original instruction. %
Validation %
instructions are %
inserted for \textit{all} relevant original instructions, %
 ensuring that errors impacting validation instructions %
 can propagate through dependency chains to comparison instructions (unless they get logically~masked,~\S\ref{subsec:symptoms}).

\subsection{\approach{} Design and Implementation}
\label{sec:llvm-implementation}

\myparagraph{Design Choices}
Intra-thread instruction checking techniques have been implemented at different levels of the software stack: as source code transformations~\cite{Hong:2010:qed,QED-difficult-bugs,ACFC}, within compiler IR~\cite{Multiple_inputs,IPAS,flowery,lu2014sdctune} or backend~\cite{Swift,nZDC,Shoestring} passes, and during post-processing of generated assembly~\cite{EDDI,CFCSS,fast-low-level24}. 
 To achieve portability %
necessary for fleet-scale deployment across  diverse microarchitectures,
ITHICA operates at the LLVM IR level; IR passes are target-agnostic and can be integrated into existing build systems without compiler %
modifications.

ITHICA %
divides checking into four %
individual
passes organized by check type~(Table~\ref{tab:transformations}), and separately implements their combinations (\AM{}, \AMD{}, \AMDB{}). 
This modularity allows independent deployment and controlled evaluation of each pass's detection contribution. 
A design distinction from prior work on soft error detection is check instruction placement~(\S\ref{sec:ithica-passes}).
There, logical masking of errors before they become externally visible~(i.e.,~at stores and branches~\cite{Swift,EDDI}) is desirable, so checks are inserted only at those points. For defects, logical masking prevents detection, so ITHICA supports block sizes down to~1, giving every instruction its own check. Block size of~1 further enables instruction localization, which is useful for characterizing permanent defects~(\S\ref{Results3}--\S\ref{Results5}), but uninformative for transient soft errors, which do not recur.

\begin{table}[!t]
\caption{ITHICA transformation passes and validation rules}
\begin{center}
\renewcommand{\arraystretch}{1.0}
\setlength{\tabcolsep}{1pt}
\scriptsize
\begin{tabular}{|l|l|l|l|l|}
\hline
\textbf{Pass} & \textbf{Target Inst} & \textbf{Original Inst} & \textbf{Validation Inst} & \textbf{Check}  \\
\hline
Arith &
\makecell[l]{Logic, binary,  bitwise, \\ vector, unary, aggregate \\\& conversion ops, \textit{GEP}, \\ \textit{icmp}, \textit{fcmp},  \textit{select,} intrin-\\ sics \& inline asm~\cite{llvm-language-ref}}
& r0 = op(s0,s1) & r1 = op(s0,s1) & eq(r0,r1) \\
\hline
\multirowcell{2}[-4pt][l]{Mem} & \makecell[l]{Load (non-atomic,\\non-volatile)} & v0 = load(a0) & v1 = load(a0) & eq(v0,v1)\\
\cline{2-5}
 & \makecell[l]{Store (non-atomic,\\non-volatile)} & store(v0,a0) & v1 = load(a0) & eq(v0,v1)\\
\hline
\multirowcell{2}[-16pt][l]{MemDiv} & \makecell[l]{Load (non-atomic,\\non-volatile)} & v0 = load(a0) &
\makecell[l]{v1 = load(a0);\\clflush(a0);\\v2 = load(a0)} &
\makecell[l]{eq(v0,v1) \&\&\\eq(v0,v2)} \\
\cline{2-5}
 & \makecell[l]{Store (non-atomic,\\non-volatile)} & store(v0,a0) &
\makecell[l]{v1 = load(a0);\\mfence();\\v2 = load(a0);\\clflush(a0);\\v3 = load(a0)} &
\makecell[l]{eq(v0,v1) \&\&\\eq(v0,v2) \&\&\\eq(v0,v3)} \\
\hline
Br & \makecell[l]{Conditional\\branch} & br(c0,t0,t1) &
\makecell[l]{src: e = c0 ? t0:t1;\\\hspace{1.8em}store(e,tmp)\\t0: \hspace{0.35em}v0 = load(tmp)\\t1: \hspace{0.35em}v1 = load(tmp)} &
\makecell[l]{src: None\\t0: \hspace{0.35em}eq(v0,t0)\\t1: \hspace{0.35em}eq(v1,t1)} \\
\hline
\end{tabular}
\label{tab:transformations}
\end{center}
\end{table}

\myparagraph{Implementation Challenges}
The first challenge, unique to ITHICA's LLVM IR implementation, is \textit{redundancy preservation}. 
Despite applying ITHICA as the final IR-level pass to minimize interference with compiler optimizations, backend optimizations can still eliminate inserted instructions. %
To prevent this without modifying the LLVM compiler backend, validation loads and stores are marked as volatile; for arithmetic instructions, which cannot be marked volatile directly, their arguments are passed through no-op volatile store-load pairs to achieve the same effect. %

The second challenge is ensuring instrumentation \textit{correctness}, which requires two conditions:
validation instructions need to be side-effect free (to preserve observable program behavior), and original and validation instructions need to yield the same output in the absence of hardware errors. %
Most %
instructions covered by \A{}
always %
meet these conditions, %
and are %
duplicated unconditionally. The exceptions are LLVM intrinsics and inline assembly, %
which are first 
checked against LLVM attributes~\cite{llvm-language-ref} %
to confirm they are side-effect free, and only then  duplicated. %
\B{} is safe by construction, as it 
inserts a validation stack store and load with no side-effects.
For \M{} and \MD{}, there are multiple original instructions with side-effects for which a validation copy  would alter program behavior---including 
atomic and volatile loads and stores, \textit{atomicrmw}, \textit{cmpxchg}, and \textit{alloca}~\cite{llvm-language-ref}%
---which are therefore excluded.
Non-atomic/non-volatile loads and stores are safely instrumented under a thread-safety assumption; permitted data races in thread-unsafe multithreaded code can cause false positives. %
For example, OpenSSL intentionally permits certain data races~\cite{openssl-manual}, so ITHICA does not instrument those functions  with \M{} and \MD{} in our experiments~(\S\ref{results}).

\section{Testing Campaign: \approach{} on Real Hardware}
\label{case-study}
To test our insight (\S\ref{key-insight}) and evaluate %
\approach{} tests 
 generated from various programs (\S\ref{sub:software-setup}), %
we conduct a large %
testing campaign (\S\ref{sub:settup}).  %

\subsection{Instrumenting Diverse Input Programs}
\label{sub:software-setup}

We use \approach{} to instrument
three input program types: (i)~industry-representative functional test programs comprising Google's \textbf{\cpucheck{} (\CC{})}~\cite{cpu-check}; 
 (ii)
industry-representative workloads comprising 
Google's \textbf{Fleetbench (\FB{})}~\cite{fleetbench,Fleetbench-paper}; and (iii) %
\textbf{libraries}~\cite{zlib,openssl,abseil,libcllvm,libcxxllvm} used by \CC{} and \FB{}.
When \approach{} generates a functional test from a program, %
we append ``\texttt{-}'' and then  ``\texttt{ITHICA}'' or the name of the applied transformation~(e.g.,~``\texttt{Arith}'') 
to the program name to get the new test name~(e.g.,~\CCithica{} or \texttt{CC-Arith}, respectively).

\looseness=-1
\myparagraph{cpu-check} \CC{} is a suite of bespoke functional tests  with final output checks~(\S\ref{sec:intro}) developed 
by Google 
for in-datacenter testing~(\S\ref{sec:background:datacentertest}). 
Test programs include: invertible computations in sequence
(e.g., encryption-decryption) %
with input-output comparisons; checksum and hash computations on random, non-deterministic data
with cross-core output comparisons; and AVX computations with vector lane output comparisons. 
Most of our experiments evaluate ITHICA tests generated from CC, since its native
checks provide a comparison~baseline,  %
and its C++
implementation allows for microarchitecture-independent LLVM
compilation (unlike SiliFuzz tests~\cite{google:silifuzz}). %

\myparagraph{Fleetbench}
\FB{} consists of C++ microbenchmarks, representing hot functions in Google datacenters, including Proto, Swissmap, Libc, Tcmalloc, Hashing, Compression, and Stl-cord, %
executed with deterministic data pre-sampled from a production distribution or randomly generated with a fixed seed. %
We exclude Compression %
due to its reliance on \textit{OpenMP}, which we could neither statically compile nor access as a system library on our servers.

\myparagraph{Libraries} %
For both
\CC{} and \FB{}, %
we instrument both top-level and library code.
For \CC{}, we apply \approach{} to \texttt{Libc}, \texttt{Libc++}, \texttt{Abseil}, \texttt{OpenSSL}, and \texttt{Zlib}. Due to tight coupling of \texttt{Libc} and \texttt{Libc++} with GCC, we instead apply \approach{} to LLVM's overlay libraries, \texttt{Llvmlibc} and \texttt{Llvmlibc++}. %
For \FB{}, we do not explicitly instrument libraries,  like we do for \CC{}, but rely on \FB{}'s build system (Bazel) to rebuild all relevant libraries~(e.g.,~Abseil) with \approach{}.

\subsection{Two-Pool Evaluation Strategy}
\label{sub:settup}

We conduct experiments on two groups of CPU servers drawn from  %
\Company{}'s fleet (consisting of many million CPUs): the \textbf{Quarantine Pool (QPool)} and the \textbf{Defective Pool~(DPool)}.

The QPool consists of more than \textbf{3,000  servers} quarantined by %
\Company{} for out-of-production testing due to suspicious behavior. They span \textbf{at least 10 microarchitectures} across two  CPU server vendors.
It is a \emph{dynamic} pool (i.e., new servers are regularly added and removed)  containing both suspect- and confirmed-defective servers. 
Servers are offlined to the QPool %
via three mechanisms: 
(1)~customer complaints about data corruption; (2) flagging by 
opportunistically-run 
 hyperscaler- and vendor-developed screening tests; and (3) forensics-based analysis of hardware and software exceptions (e.g., fail-stop and fail-slow errors including machine check exceptions, kernel or user crashes,  invariant fails in user code). 

The DPool consists of 20 confirmed-defective servers, originally part of the larger QPool, that we set aside for dedicated testing with ITHICA.
All 20 have failed at least one functional test developed either by %
\Company{} or by the %
CPU vendor %
and exhibit diversity in terms of microarchitecture %
and age~(Table~\ref{tab:dpool-characteristics}). %
Across all our experiments, no functional test~(i.e.,~neither \approach{} nor \CC{}) detects any error on six of them.
Thus, we report results for the remaining \textbf{14 servers}.

\begin{table}[t!]
\caption{%
DPool Characteristics  %
} 
\centering
\setlength{\tabcolsep}{1.7pt}
\renewcommand{\arraystretch}{0.90}
\renewcommand{\arraystretch}{1.0}
\begin{tabular}{@{}l*{14}{c}@{}}
\toprule
 & \textbf{D1} & \textbf{D2} & \textbf{D3} & \textbf{D4} & \textbf{D5} & \textbf{D6} & \textbf{D7} & \textbf{D8} & \textbf{D9} & \textbf{D10} & \textbf{D11} & \textbf{D12} & \textbf{D13} & \textbf{D14} \\ \midrule
\textbf{$\mu$}\textbf{arch}          & u1 & u2 & u3 & u4 & u5 & u5 & u2 & u6 & u1 & u4 & u3 & u5 & u1 & u2 \\
\textbf{Age (months)}           & 19 &  6 & 34 & 16 & 45 & 38 & 81 & 19 & 24 & 11 & 11 & 42 & 10 & 64 \\
\bottomrule
\end{tabular}
\label{tab:dpool-characteristics}
\end{table}

\myparagraph{Benefits of Two-Pool Evaluation} The DPool and QPool serve distinct roles.
The DPool provides a controlled environment with confirmed-defective servers, enabling us to %
evaluate our %
key insight~(\S\ref{key-insight}). Its static nature and our exclusive access to it ensure sufficient runtime to evaluate all \approach{} transformations and configurations: we collect \textbf{over 2,000 hours of testing time per DPool server}, exceeding the allocated time of previous detailed studies by 1--2 orders of magnitude~\cite{sevi-meta,Wang:alibaba:sdc,alibaba-new}.
In the QPool, ITHICA tests integrate into \Company{}'s standard out-of-production testing pipeline alongside other industrial tests, enabling large-scale evaluation in a real-world in-datacenter setting.

\section{Experiments Overview}
\label{Experiments}

The experimental evaluation in this paper is partitioned %
into two~subsections: testing on the DPool~(\S\ref{sec:experiment1}) and testing on the QPool~(\S\ref{sec:experiment2}). %
Across our experiments, we compare the efficacy of \approach{} tests to \CC{} (our baseline) using four main metrics, defined in the first four rows of Table~\ref{tab:metrics}. The results are interpreted in detail in~\S\ref{results}.

\begin{figure*}[t!]
    \centering
    \includegraphics[width=1.0\linewidth]{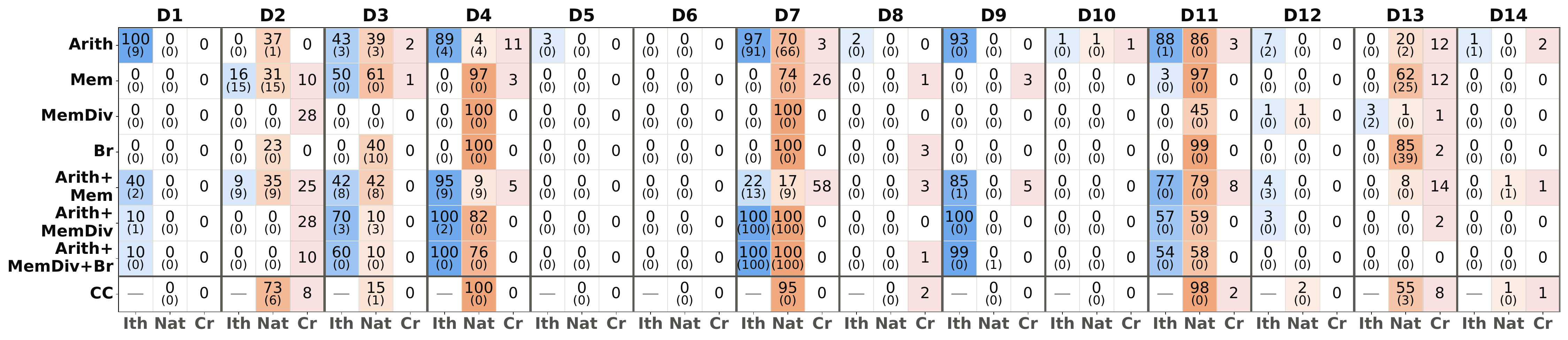}
    \caption{\texttt{EDR} of %
    \CCithica{} %
    (with block size and interleaving of~1) and \CC{} in DPool.  %
    Each triplet reports results for \ITHICA{} (\textit{Ith}), \Native{} (\textit{Nat}), and %
    crashes  with no detection (\textit{Cr}).
    The subset of runs that crashed after %
    detection is shown in parentheses.
    \texttt{D6} is uniquely detected by \A{} for interleaving of~8~(\S\ref{Results3}).
    }
    \label{fig:heatmap} 
\end{figure*}

 \begin{figure}[t!]
    \centering
        \centering
        \includegraphics[trim={0 0 0 0.38cm},clip,width=\linewidth]{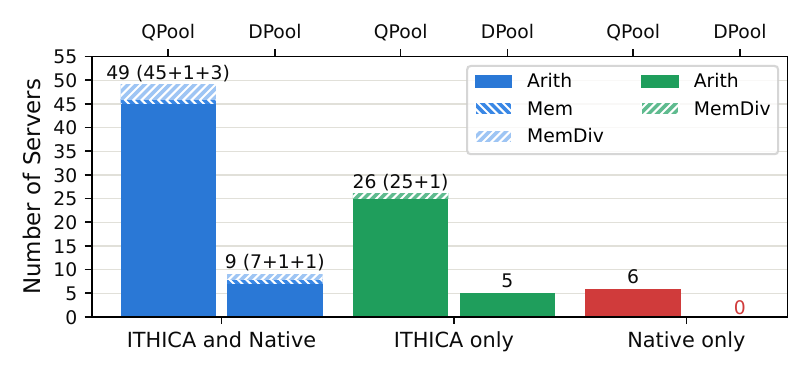}
        \caption{
        Server detections across all \CCithica{} runs in both pools. %
        }
        \label{fig:qpool-new}
\end{figure}

\subsection{Testing on the DPool}
\label{sec:experiment1}
First, to compare \CCithica{} to \CC{}, we run 
\textbf{eight functional tests} on each DPool server. Seven are \CCithica{} tests obtained by applying its four main transformations (\S\ref{sec:ithica-passes}) and three combinations thereof (\AM{}, \AMD{}, \AMDB{}) on \CC{}, all with block size and interleaving set to 1. The eighth is the original \CC{} (i.e., without ITHICA instrumentation) for comparison. 
Each test runs 100 times per server (\textbf{100 runs})  each lasting \textbf{one hour} unless interrupted by a crash. Between runs, the server is rebooted. 

Within an ITHICA test, we distinguish between \Native{} checks (checks in the original program, i.e., \CC{}'s final output checks) and \ITHICA{} checks (inserted by \approach{}).  
If either check type detects an error before a crash, the detection counts. 
\CC{} contains only \Native{} checks; \CCithica{} contains both. 
Each run executes hundreds of thousands of rounds of the \CCithica{} program (and more rounds for \CC{}), each with different randomly-generated inputs, yielding hundreds of thousands of inputs per run.

Fig.~\ref{fig:heatmap} shows \texttt{EDR} (Table~\ref{tab:metrics}) for the seven \CCithica{} tests and \CC{} across all 14 DPool servers (D1–D14). Parenthetical values show the subset of detections followed by crashes; the~\textit{Cr} column shows crashes without any other detection.

\looseness=-1
To compare different \approach{} interleaving and block size configurations (\S\ref{sec:configuring}) for the same program, we conduct two other experiments.
First, fixing block size to 1, we \textbf{sweep interleavings} of 1, 2, 4, 8, and \texttt{max} (basic block length). Second, fixing interleaving to 1, we \textbf{sweep block sizes} of 1, 2, 4, 8, and \texttt{dep}, which 
inserts comparisons at the end of each instruction's dependency chain. 
We evaluate each configuration with \textbf{100 one-hour runs}.
Fig.~\ref{fig:mapbs} shows the results of these two experiments. %

Finally, we evaluate ITHICA test content generated from two types of non-test programs, using \A{} %
with block size and interleaving of 1, since this configuration yielded the most detections on the DPool for \CCithica{} (Fig.~\ref{fig:heatmap},~\S\ref{Results1}): \textbf{five libraries} used by CC 
and \textbf{six \FB{} workloads} (\S\ref{sub:software-setup}). %
We evaluate each \texttt{FB-ITHICA} binary for \textbf{100 one-hour runs}.

\begin{table}[t!]
\caption{Evaluation metrics -- each run spans one hour}
\raggedright
\setlength{\tabcolsep}{1pt}
\renewcommand{\arraystretch}{1.0}
 \setlength{\aboverulesep}{1pt}  
  \setlength{\belowrulesep}{1pt}  
\begin{tabular}{@{}l| l@{}}
\toprule
 \textbf{Metric} & \textbf{Description} \\ 
 \midrule
Server Detections &  \# servers with error detections\\
Error Detection Rate (EDR) &  \# runs with error detections / \# all runs (\%)\\
Error Frequency (EF) &  \makecell[l]{\# error detections / \# all runs} \\
Time to Detection (TTD) &  \makecell[l]{time from test start to first error} \\
\midrule
PC Sensitivity \textit{(per opcode)} &  \makecell[l]{\# failing static PCs / \# all static PCs (\%)} \\
BB Sensitivity \textit{(per opcode)} &  \makecell[l]{\# failing basic blocks (BBs) / \# all BBs (\%)} \\
Input Breadth \textit{(per opcode)} &  \makecell{\# unique failing inputs / \# all failing inputs (\%)} \\
\bottomrule
\end{tabular}
\label{tab:metrics}
\end{table}

\subsection{Testing on the QPool}
\label{sec:experiment2}

Due to limited available execution time in the QPool compared to the DPool, we primarily run \textbf{\texttt{CC-Arith}} 
with block size and interleaving  of 1, %
again due to its success for \CCithica{} in the DPool.
The total execution time accumulated per server varies, ranging from \textbf{20 to 100 hours} due to the presence of other tests in the pipeline and the dynamic~nature~of~the~QPool.

We secondarily run \textbf{\texttt{CC-Mem}, \texttt{CC-MemDiv}, and \texttt{CC-Br}} with block size and interleaving of 1,
reporting \textit{only their unique detections} compared to \texttt{CC-Arith}
due to their shorter runtime~(at least \textbf{20 hours}). The results %
are summarized in Fig.~\ref{fig:qpool-new}, 
alongside the aggregate DPool results.  Our baseline remains \Native{} checks within the same \CCithica{} binaries. To ensure reliable analysis, these results exclude servers~with~less~than~20~hours~per~test. 

Finally, as in the DPool, we evaluate \CC{}'s \textbf{five libraries} and the \textbf{six \FB{} workloads} transformed with \A{} with block size and interleaving of 1 for \textbf{20 to 100 hours}.

\section{Results}
\label{results}

We present ten key findings from the experiments in~\S\ref{sec:experiment1} and~\S\ref{sec:experiment2}, along with follow-up experiments for further analysis.  
We first establish ITHICA's effectiveness 
for defect detection~(\S\ref{Results1},~\S\ref{Results2}). We then leverage ITHICA's instruction-level visibility to  characterize defect behavior and the factors governing  detection~(\S\ref{Results3}--\S\ref{Results5}),  contrasting our findings with prior work's conclusions~(\S\ref{Results4},~\S\ref{Results5}).

\subsection{Defects Manifest as Inconsistent Errors}
\label{Results1}
First, consider Fig.~\ref{fig:heatmap}, produced from running seven \CCithica{} tests and \CC{} in the DPool (\S\ref{sec:experiment1}). \obs{}~The figure shows that  \CCithica{} tests collectively detect %
all 14 DPool servers. 
Among the four base transformations, \A{} is the most effective, detecting 11 out of 14 servers. 
This result %
may reflect survivorship bias~(\S\ref{limitations}) in the tests used~to flag servers for offline testing (\S\ref{sub:settup}).
\B{} does not detect~errors in the DPool but flags non-unique servers in the QPool~(\S\ref{sec:experiment2}). %
\obs{} Four servers (D1, D5, D8, D9) are uniquely detected by \ITHICA{} checks in \CCithica{} tests, while missed by \Native{} checks in the same binaries as well as by \CC{}.
D6 is also uniquely detected by \A{} for interleaving of 8~(\S\ref{Results3}).

Next, consider Fig.~\ref{fig:qpool-new}, produced from running four/seven \texttt{CC-ITHICA} tests in the QPool/DPool (\S\ref{sec:experiment2}/\S\ref{sec:experiment1}).  
Of the~49 QPool servers detected by both \ITHICA{} and \Native{} checks in \CCithica{} tests, 45 were flagged by \A{}, 1 uniquely by \M{}, and 3 uniquely by \MD{}.
\obs{} An additional 26 QPool servers were only captured by \ITHICA{}; 25 by \A{} and 1 uniquely by \MD{}. 
Only six servers were detected by \Native{} but not  \ITHICA{}. 
Two likely explanations are gaps in ITHICA's instrumentation and %
errors that manifest consistently, %
discussed in~\S\ref{limitations}.
Across both pools, \CCithica{} tests detect~89 total servers~(75/14 in the QPool/DPool)---\textbf{39\% more than \Native{} checks} in the same binaries. 
Finally, \approach{} tests derived from \FB{} (\S\ref{sec:experiment1}, \S\ref{sec:experiment2}) detect %
\textbf{11 additional servers} in the QPool (\S\ref{Results3}),~resulting in \textbf{100 servers} ($89+11$) detected by \approach{} across both pools.

\begin{table}[t!]
\footnotesize
\caption{Decomposition of  instructions exhibiting errors
  in DPool %
  }
 \centering
\renewcommand{\arraystretch}{1.0}  
  \setlength{\tabcolsep}{7pt}
 \setlength{\aboverulesep}{2pt}  
  \setlength{\belowrulesep}{2pt}  
  \begin{tabular}{c|c|c|c}
    \toprule
    \makecell{\textbf{Instruction with Error}} &
    \makecell{\textbf{Original}} &
    \makecell{\textbf{Validation}} &
    \makecell{\textbf{Both}} \\
    \midrule
    \makecell{\% across all DPool runs} & 31.3\% & 44.7\% & 24.0\% \\ 
    \bottomrule
  \end{tabular}
  \label{tab:wrong}
\end{table}

\looseness=-1
Notably, inconsistent errors are not limited to cases where only one of the two instruction instances is affected by a defect: ITHICA can detect errors even when both are. 
Table~\ref{tab:wrong} shows  which of the two instructions (original or validation) exhibits an error when \approach{} detects one across DPool runs. \obs{}~In~24\% of cases, \textit{both} produce \textit{incorrect but different} outputs. %
These cases are particularly informative because they %
constrain the possible explanations for the inconsistency: since both instances are incorrect, both were likely affected by 
the same defective %
hardware component, %
yet still produced different outputs. %
In contrast, when only one instruction is incorrect, we cannot determine whether they interacted with the same %
component or %
different ones (i.e., one defective and one not).

\begin{finding}
\textbf{Finding 1:} 
Defects, despite being permanent faults,  manifest as inconsistent errors in nearly all servers flagged as defective in our testing campaign (Obs.~1--4). 
\end{finding}

For the Fig.~\ref{fig:qpool-new} experiment, Table~\ref{tab:stats} compares  \ITHICA{} and \Native{} checks on several other metrics (Table~\ref{tab:metrics}).
\obs{}~Across 58 servers detected by both \ITHICA{} and \Native{}~(49/9 in the QPool/DPool), %
  ITHICA achieves, on average,  1.78$\times$ \texttt{EDR} improvement, 7.51$\times$ \texttt{EF} improvement and~0.68$\times$ \texttt{TTD} improvement (1.47$\times$ faster).

\begin{finding}
\textbf{Finding 2:} 
\looseness=-1
ITHICA %
checks outperform final output checks used by 
hyperscaler tests across all evaluated metrics~(Obs.~2,~3,~5).
\end{finding}

\begin{figure}[t!]
\centering
\includegraphics[width=1.01\linewidth]{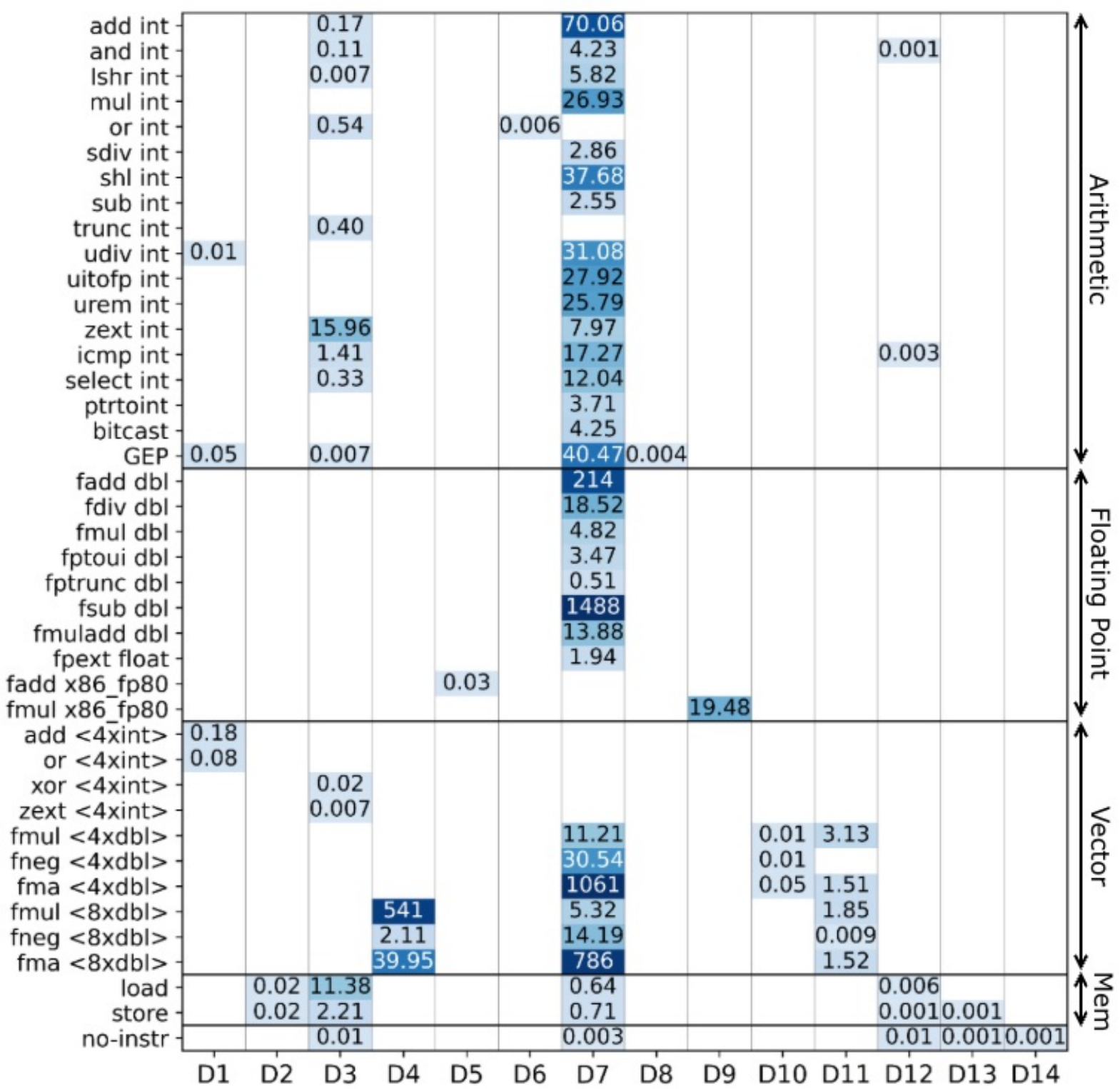}
\caption{Distribution of %
instructions where \approach{} detects errors, with \texttt{EF} calculated %
across all \CCithica{} runs in DPool. %
}
\label{fig:instructions}
\end{figure}

\looseness=-1
Fig.~\ref{fig:instructions} shows the distribution of failing instructions %
and  their average %
\texttt{EF} across all \CCithica{} tests run in the DPool with a block size of 1 and an interleaving of 1, except  D6, which is uniquely detected at interleaving of 8 (\S\ref{Results3}).
A ``failing instruction'' denotes one that exhibits an incorrect output; it does not imply a particular defective hardware unit~(\S\ref{Results5}). 
Cases where no specific instruction is identified (\textit{no-instr}) can result from an error in an instrumentation instruction itself~(e.g.,~\texttt{icmp}), or a crash before logging completes.  

\obs{} The data reveals a diverse range of affected instructions across servers, spanning arithmetic, floating-point, vector, and memory instructions. Average error frequencies also vary drastically, from less than one per run to over a thousand. Both observations reflect the fact that defects 
impact different physical regions of each chip in different ways~(\S\ref{subsubsec:permanent}).

While ITHICA can detect inconsistent errors %
regardless of their source, the observed error 
characteristics  %
in our experiments narrow plausible %
causes. \obs{} Their frequency and repeatability~(Fig.~\ref{fig:instructions}) make transient faults less plausible; their device-specific~(Fig.~\ref{fig:instructions}) yet microarchitecture-agnostic~(Table~\ref{tab:dpool-characteristics}) nature points away from design bugs; their presence across devices~of~varying ages~(Table~\ref{tab:dpool-characteristics}) challenges aging effects as a sole cause.~Taken together, these observations %
are consistent with defects as the likely cause~(\S\ref{sources}). 

\begin{table}[t!]
  \caption{Comparison of \ITHICA{} and \Native{} detection metrics (\texttt{EDR}, \texttt{EF}, \texttt{TTD}) for 58 servers detected by both checks 
}
\footnotesize
\centering
\renewcommand{\arraystretch}{1.0}
\setlength{\aboverulesep}{1pt}  
\setlength{\belowrulesep}{1pt}  
\setlength{\tabcolsep}{2pt}  %
  \definecolor{lightyellow}{RGB}{255, 255, 220} %
  \resizebox{1\linewidth}{!}{
  \begin{tabular}{c|>{\columncolor{lightyellow}}c|c|c|c|c}
    \toprule
    \makecell{\textbf{Ratio} \textbf{\ITHICA{}}/\textbf{\Native{}}} & \makecell{\textbf{Geom.} \textbf{Mean}} & \makecell{\textbf{Arith.} \textbf{Mean}} & \textbf{Median} & \textbf{Min} & \textbf{Max} \\
    \midrule
     \makecell{%
     Error Detection Rate (\texttt{EDR})} &  1.78 & 4.50  & 1.17 & 0.52 & 100.00 \\ 
     \midrule
     \makecell{Error Frequency (\texttt{EF})} & 7.51 & 130.15 & 3.44  & 0.02 & 5396.00 \\
    \midrule
    \makecell{Time to Detection (\texttt{TTD})} & 0.68 & 0.77 & 0.56  & 0.33 & 1.72
    \\ 
    \specialrule{\heavyrulewidth}{0pt}{0pt}
  \end{tabular}
  }
  \label{tab:stats}
\end{table}

\begin{finding}
\textbf{Finding 3:}
ITHICA-detected errors in our server pools are most consistent with defects as the underlying cause~(Obs.~6,~7).
\end{finding}

\implication{Implications for Chip-Users:}{Intra-thread, instruction-level checking for inconsistent errors is highly effective for defect~detection.  This finding---not 
leveraged in any 
existing~functional test for defects~\cite{intel:opendcdiag,cpu-check,google:silifuzz,harpocrates,harpocrates_plusplus,Wang:alibaba:sdc,alibaba-new,pindrop,sevi-meta,Dixit:2021:sdc-at-scale,dixit:sdc-detecting}---enables  %
 the use of arbitrary programs as tests, removing~bias toward those compatible with existing error
 checking techniques~(\S\ref{sec:intro}).}

\subsection{\approach{} Paves the Way for Online Testing}
\label{Results2}

The permanent nature of defects 
has a direct %
implication for ITHICA instrumentation.
For defect detection (i.e., identifying whether a server is defective), catching \textit{any one} error per server suffices, unlike soft error detection which requires catching \textit{every} corruption that affects an application's output. 
This distinction enables ITHICA instrumentation to be reduced, trading \texttt{EDR} 
for lower overhead, while 
maintaining high defect detection coverage (i.e., number of defective servers detected).
Notably, for error characterization, fine-grained instrumentation at every instruction remains necessary~(\S\ref{Results5}).

\begin{figure*}[t!]
    \centering
    \includegraphics[trim={0 0 0 2.05cm},clip,width=\linewidth]{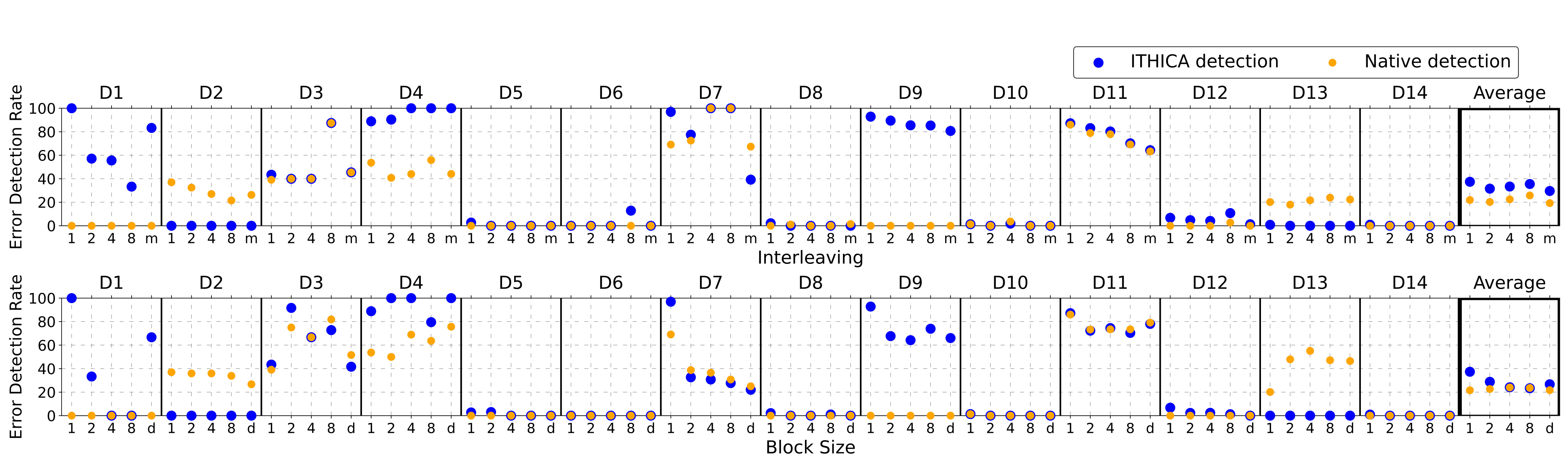}
    \caption{ \label{fig:mapbs} 
Impact of interleaving and block size on \texttt{EDR} %
for \texttt{CC-Arith} in DPool. %
Top row: %
effect of varying interleaving (m=\textit{max}, length of the basic block). 
Bottom row: %
effect of varying block size (d=\textit{dep}, length of instruction dependency chain). The rightmost panels show  average \texttt{EDR} across all
servers. 
}
\end{figure*}

Two %
strategies can be used, separately or in combination, for reduced ITHICA
instrumentation: \textit{sampling in time} (i.e., choosing the check insertion frequency in checked program regions) and \textit{sampling in space} (i.e., choosing which program regions are checked). %
In this paper, we investigate a sampling in time strategy: %
varying \textit{block size}~(\S\ref{sec:configuring}). %
Fig.~\ref{fig:mapbs} shows that  a
block size of 1 performs best on average, likely due to 
 reduced logical error masking.
 \obs{} However, larger block sizes demonstrate comparable effectiveness: all DPool servers with reproducible  (i.e., more than one) detections at block size 1 (all except for D10 and D14)  are also detected at larger block sizes. This indicates that checking frequency can be relaxed without significantly compromising the number of detected servers. 
 The runtime overhead of \A{} decreases from an average of~2.22$\times$ at block size~1 to 1.56$\times$ at block size 8~(Table~\ref{tab:overheads}).

\begin{finding}
\textbf{Finding 4:} 
ITHICA instrumentation and runtime overhead can be reduced at high defect detection coverage (Obs. 8).
\end{finding}

\implication{Implications for Chip-Users:}{The permanent nature of defects can be exploited for relaxed instrumentation. ITHICA's fine-grained, flexible checking, easily applied to  new programs and tunable in frequency and scope, can support a wide variety of relaxation strategies. %
Establishing ITHICA's effectiveness for defects is the 
first step toward low-overhead online  testing---important future work.}

\subsection{Execution Context %
Drives Defect Detection}
\label{Results3}

ITHICA error detections appear sensitive to two mechanisms for varying execution context. First, the instruction sequence preceding an %
 opcode instance determines whether it executes %
in a context where it  \textit{can} exhibit an inconsistent
error. Second,  execution context variation via natural and~proactive diversity~(\S\ref{execution-context-diversity}) between an original instruction with that opcode and its validation instruction determines whether~it~\textit{does}.

\textit{\textbf{Sensitivity to preceding instruction sequence.}} First, we examine how preceding instruction sequences influence ITHICA error detections across different instances of the same opcode within the same test.
For each opcode-server combination that ITHICA detects an error on, %
we analyze what fraction of static program counters (PCs) executing the  opcode exhibit errors %
to calculate \textbf{PC Sensitivity} (Table~\ref{tab:metrics}). Opcodes with only one PC or one error are omitted from the table.  
\obs{} For most failing opcodes, PC Sensitivity is very low (often $<$1\%, Table~\ref{tab:pc_input_sensitivity}), revealing that the same opcode behaves correctly in the vast majority of its program locations. The same is true for basic block (BB) Sensitivity. A notable exception is vector double instructions, which exhibit high PC and BB Sensitivity. However, for each of these opcodes, all PCs are contained within %
a \textit{single basic block}, %
and therefore likely execute under a similar execution context. Similarly, other opcodes with relatively high PC Sensitivity---namely floating-point and vector integer instructions---also exhibit high BB Sensitivity, with failing PCs concentrated within one of only two basic blocks containing them~(Table~\ref{tab:pc_input_sensitivity}).

\begin{table}[t!]
\caption{Performance overhead and binary size increase (in~$\times$) of ITHICA %
  compared to \CC{}, across DPool microarchitectures}
  \centering
  \renewcommand{\arraystretch}{0.7}
  \setlength{\aboverulesep}{1pt}
  \setlength{\belowrulesep}{1pt}
  \setlength{\tabcolsep}{4pt}
  \begin{tabular}{p{0.1cm} C{0.9cm} ccccc ccc}
    \toprule
    & \multirow{2}{*}{\makecell{\textbf{ITHICA}\\\textbf{Pass}}}
      & \multicolumn{5}{c}{\textbf{\A{} (Block Size)}}
      & \multirow{2}{*}{\textbf{\M{}}} & \multirow{2}{*}{\textbf{\MD{}}} & \multirow{2}{*}{\textbf{\B{}}} \\
    \cmidrule(lr){3-7}
    & & (1) & (2) & (4) & (8) & (dep) & & & \\
    \midrule
    \multirow{6}{*}{\rotatebox[origin=c]{90}{\textbf{Performance}}}
    & u1 & 2.26 & 2.06 & 1.88 & 1.71 & 1.85 & 1.13 & 29.36 & 1.11 \\
    & u2 & 2.23 & 1.88 & 1.66 & 1.58 & 1.43 & 1.10 & 34.89 & 1.10 \\
    & u3 & 2.28 & 1.93 & 1.75 & 1.62 & 1.82 & 1.23 & 29.63 & 1.18 \\
    & u4 & 2.30 & 1.70 & 1.56 & 1.49 & 1.63 & 1.11 & 50.17 & 1.09 \\
    & u5 & 2.07 & 1.70 & 1.53 & 1.42 & 1.44 & 1.09 & 48.49 & 1.08 \\
    & u6 & 2.17 & 1.80 & 1.61 & 1.54 & 1.69 & 1.07 & 57.23 & 1.05 \\
    & \textbf{Avg} &  2.22  & 1.85  &  1.67   &  1.56   &  1.64   &  1.12   &  41.63   &  1.10   \\
    \midrule
    \multicolumn{2}{c}{\textbf{Binary Size}} & 8.47 & 5.88 & 4.59 & 4.11 & 6.11 & 3.06 & 13.12 & 1.24 \\
    \bottomrule
  \end{tabular}
  \label{tab:overheads}
\end{table}

Next, we show that applying different ITHICA transformations that check at least one common opcode to the same program varies the preceding instruction sequences of those opcodes, similarly affecting detections.
\obs{}~Per Fig.~\ref{fig:heatmap}, combined passes (\AM{}, \AMD{}, \AMDB{}) have the benefit of detecting %
errors across multiple instruction types (e.g., both arithmetic and memory instructions for D12), 
but sometimes have lower \texttt{EDR} or entirely miss servers detected by individual passes. This can be attributed to %
the additional instrumentation instructions changing the instruction sequence compared to individual passes, and thus disrupting the execution context needed~for~detection.

\begin{table*}[t!]
\caption{ 
PC Sensitivity, BB Sensitivity and Input Breadth per opcode and per DPool server, across  \CCithica{} runs %
}
\centering
\setlength{\tabcolsep}{1.0pt}
\definecolor{lightgreen}{RGB}{220, 250, 220}
\definecolor{lightgray}{gray}{0.95}
\definecolor{warmgray}{RGB}{232,228,220}
\renewcommand{\arraystretch}{0.40}
\setlength{\aboverulesep}{1pt}
\setlength{\belowrulesep}{1pt}
\scalebox{0.815}{
\begin{tabular}{l|l|c|ccc|ccc|ccc}
\toprule
\textbf{Category} &
\multicolumn{1}{c|}{\makecell{\textbf{Failing} \\ \textbf{Opcode}}} &
\makecell{\textbf{Affected} \\ \textbf{Servers}} &
\makecell{\textbf{\# Total} \\ \textbf{PCs}} &
\makecell{\textbf{\# Failing} \\ \textbf{PCs}} &
\makecell{\textbf{PC Sensitivity} \\ \textbf{Geom. Mean (Std)}} &
\makecell{\textbf{\# Total} \\ \textbf{BBs}} &
\makecell{\textbf{\# Failing} \\ \textbf{BBs}} &
\makecell{\textbf{BB Sensitivity} \\ \textbf{Geom. Mean (Std)}} &
\makecell{\textbf{\# Total} \\ \textbf{Failing Inputs}} &
\makecell{\textbf{\# Unique} \\ \textbf{Failing Inputs}} &
\makecell{\textbf{Input Breadth} \\ \textbf{Geom. Mean (Std)}} \\
\midrule
\cellcolor{white} & add int & D3 & 1,202 & 5 & 0.42\% (0.00) & 664 & 4 & 0.60\% (0.00) & 22 & 7 & 31.82\% (0.00) \\
\rowcolor{lightgray} \cellcolor{white} & and int & D3 / D12 & 748 & 3 / 1 & 0.23\% (0.13) & 634 & 3 / 1 & 0.27\% (0.00) & 2 / 1 & 2 / 1 & 100.00\% (0.00) \\
\cellcolor{white} & lshr int & D3 & 424 & 1 & 0.24\% (0.00) & 203 & 1 & 0.49\% (0.00) & 1 & 1 & 100.00\% (0.00) \\
\rowcolor{lightgray} \cellcolor{white} & or int & D3 / D6 & 180 & 7 / 1 & 1.47\% (1.67) & 145 & 7 / 1 & 1.82\% (2.07) & 776 / 9 & 34 / 2 & 9.87\% (8.92) \\
\cellcolor{white} & trunc int & D3 & 492 & 5  & 1.02\% (0.00) & 240 & 5 & 2.08\% (0.00) & 6 & 4 & 66.67\% (0.00) \\
\rowcolor{lightgray} \cellcolor{white} & udiv int & D1 & 36 & 1 & 2.78\% (0.00) & 25 & 1 & 4.00\% (0.00) & 1 & 1 & 100.00\% (0.00) \\
\cellcolor{white} & zext int & D3 & 809 & 46 & 5.69\% (0.00) & 489 & 32 & 6.54\% (0.00) & 19{.}71K & 217 & 1.10\% (0.00) \\
\rowcolor{lightgray} \cellcolor{white}\multirow{-12}{*}[4pt]{\makecell[l]{Integer\\ Arithmetic}} & getelementptr & D1 / D3 / D8 & 5,568 & 6 / 1 / 1 & 0.03\% (0.04) & 2,412 & 6 / 1 / 1 & 0.08\% (0.10) & 3 / 1 / 4 & 3 / 1 / 1 & 63.00\% (35.36) \\
\midrule
\cellcolor{white} & icmp int & D3 / D12 & 3,968 & 12 / 1 & 0.09\% (0.14) & 3,552 & 12 / 1 & 0.10\% (0.15) & 17 / 1 & 4 / 1 & 48.51\% (38.24) \\
\rowcolor{lightgray} \cellcolor{white}\multirow{-3.0}{*}[1pt]{\makecell[l]{Comparisons}} & select int & D3 & 617 & 7 & 1.13\% (0.00) & 419 & 7 & 1.67\% (0.00) & 84 & 4 & 4.76\% (0.00) \\
\midrule
\cellcolor{white} & fadd x86\_fp80 & D5 & 4 & 1 & 25.00\% (0.00) & 2 & 1 & 50.00\% (0.00) & 36 & 6 & 16.67\% (0.00) \\
\rowcolor{lightgray} \cellcolor{white}\multirow{-2.7}{*}[1pt]{\makecell[l]{FP Arithm}} & fmul x86\_fp80 & D9 & 4 & 1 & 25.00\% (0.00) & 2 & 1 & 50.00\% (0.00) & 17.92M & 1{.}33K & 0.01\% (0.00) \\
\midrule
\cellcolor{white} & add $<$4xint$>$ & D1 & 11 & 2 & 18.18\% (0.00) &2&1&50.00\% (0.00) & 1 & 1 & 100.00\% (0.00) \\
\rowcolor{lightgray} \cellcolor{white}\multirow{-4.1}{*}{\makecell[l]{Vector\\$<$4xint$>$}} & or $<$4xint$>$ & D1 & 2 & 1 & 50.00\% (0.00) &2&1&50.00\% (0.00) & 16 & 4 & 25.00\% (0.00) \\
\midrule
\cellcolor{white} & fmul $<$4xdbl$>$ %
& D10 / D11 & 8 & 3 / 8 & 61.24\% (31.25) &1&1&100.00\% (0.00) & 6 / 740{.}92K & 4 / 1{.}57K & 3.76\% (33.23) \\
\rowcolor{lightgray} \cellcolor{white} & fneg $<$4xdbl$>$ %
& D10 & 8 & 5 & 62.50\% (0.00) &1&1&100.00\% (0.00) & 2 & 2 & 100.00\% (0.00) \\
\cellcolor{white}\multirow{-6.0}{*}[-1.5pt]{\makecell[l]{Vector\\$<$4xdbl$>$}} & fma $<$4xdbl$>$ %
& D10 / D11 & 8 & 7 / 5 & 73.95\% (12.50) &1&1&100.00\% (0.00) & 77 / 303{.}20K & 19 / 803 & 2.56\% (12.21) \\
\midrule
\rowcolor{lightgray} \cellcolor{white} & fmul $<$8xdbl$>$ %
& D4 / D11 & 8 & 8 / 8  & 100.00\% (0.00) &1&1&100.00\% (0.00) & 20.10B / 169{.}20K & 342{.}02K / 838 & 0.03\% (0.25) \\
\cellcolor{white} & fneg $<$8xdbl$>$ %
& D4 / D11 & 8 & 8 / 1 & 35.36\% (43.75) &1&1&100.00\% (0.00) & 15{.}55K / 1 & 280 / 1 & 13.42\% (49.10) \\
\rowcolor{lightgray} \cellcolor{white}\multirow{-6.0}{*}[-1.5pt]{\makecell[l]{Vector\\$<$8xdbl$>$}} & fma $<$8xdbl$>$ %
& D4 / D11 & 8 & 8 / 4 & 70.71\% (25.00) &1&1&100.00\% (0.00) & 313.97M / 261{.}97K & 40{.}02K / 810 & 0.06\% (0.15) \\
\midrule
\cellcolor{white} & load & D2 / D3 / D12 & 4,481 & 1 / 66 / 2 & 0.11\% (0.68) & 2,754 & 1 / 37 / 2 & 0.15\% (0.61) & - / 4{.}06K / 1 & - / 29 / 1 & 8.45\% (49.64) \\
\rowcolor{lightgray} \cellcolor{white}\multirow{-3.1}{*}{\makecell[l]{Memory}} & store & D2 / D3 / D12 / D13 & 3,015 & 1 / 19 / 1 / 1 & 0.07\% (0.26) &1,367 & 1 / 14 / 1 / 1 & 0.14\% (0.41)& 1 / 266 / 1 / 1 & 1 / 8 / 1 / 1 & 41.64\% (42.00) \\
\bottomrule
\end{tabular}
}
\label{tab:pc_input_sensitivity}
\end{table*}

Finally, instruction sequences preceding checked opcodes vary across input programs, likewise influencing detections.
Table~\ref{tab:total-libraries} shows detected servers across
the top-level \CC{} code in \CCithica{} tests, its library code, and \FB{}. Since \CC{} comprises multiple subtests (\S\ref{sub:software-setup}), we separate results per subtest. %
For each, we report total server detections 
and unique server detections compared to other tests or subtests.

Among \CC{} subtests, \texttt{Avx}  detects the most unique servers. This is unsurprising since it is the only test exercising large~(4$\times$,~8$\times$) vector operations. 
Among the instrumented libraries, Zlib detects the most servers (23 total). \obs{}~Of~the~31 servers missed by \Native{} in \CCithica{}~(Fig. \ref{fig:qpool-new}), \textbf{eight are detected exclusively by \ITHICA{} within library code}: six in Zlib only, one in OpenSSL only, and one in both Zlib and OpenSSL.
This demonstrates that 
even  %
if \Native{} checks were manually constructed to be more fine-grained in the top-level code, 
they could still miss errors that manifest in library code and are masked before reaching the library function's~output. 

\obs{} \FB{} tests (\S\ref{sec:experiment1}/\S\ref{sec:experiment2}) detect 24 servers (20/4 in the QPool/DPool), 
including \textbf{11 unique QPool servers} (not in Fig.~\ref{fig:qpool-new}) missed by both top-level \CC{} (both \ITHICA{} and \Native{}) %
and libraries, for a \textbf{total of 100 servers detected by ITHICA}. Importantly, none of the \FB{} programs could be used as tests without ITHICA, as they lack built-in checkable outputs. 

\begin{finding}
\textbf{Finding 5:} 
Whether an instruction exhibits a defect-induced  
error depends on whether it executes in a vulnerable context 
shaped by its preceding
instruction sequence
(Obs. 9--12). 
\end{finding}

\textbf{\textit{Sensitivity to natural or proactive diversity.}}
\obs{}~Fig.~\ref{fig:mapbs} shows that natural diversity
with an interleaving of 1 (where  each validation instruction immediately follows its original) is sufficient to expose inconsistent errors for most servers in the DPool. 
However, larger interleavings %
further improve \texttt{EDR} for some servers %
(e.g., D3, D4).  Notably, D6 is exclusively detected at an interleaving of 8. %

Beyond natural diversity, 
proactive diversity may be necessary to expose certain defects. \obs{} As shown in Fig.~\ref{fig:heatmap}, D13 is uniquely detected by \MD{}; all other ITHICA transformations miss it. Errors are localized to the third validation load %
that checks an original store (and follows a \texttt{clflush}) while the previous two loads yield correct results.
One additional server in the QPool is similarly detected exclusively by \MD{} (Fig.~\ref{fig:qpool-new}). This suggests that the \texttt{clflush} and \texttt{mfence} instructions inserted by \MD{} perturb the server's execution context between original and validation memory instructions sufficiently to expose an inconsistent error.

These observations show that  \emph{natural diversity}~(\S\ref{execution-context-diversity})~from modest interleavings is generally sufficient to expose inconsistencies in arithmetic instructions, possibly because their relevant execution context is largely core-side. \emph{Proactive diversity}~(\S\ref{execution-context-diversity})
appears necessary %
for certain defects affecting memory instructions, %
possibly due to their
relevant execution context spanning core and uncore components.

\begin{finding}
\textbf{Finding 6:} 
Detection of defect-induced inconsistent errors depends on (typically modest) execution context variation between original and validation instructions~(Obs.~13,~14).
\end{finding}

\implication{Implications for Chip-Users:}{The key to 
executing instructions in diverse microarchitectural and electrical state contexts with ISA-level control is varying this context \textbf{implicitly}, by
(a) transforming arbitrary programs into tests, 
(b) varying and combining ITHICA transformations to alter the surrounding instruction mix within a test and (c) varying interleaving to exploit natural diversity within a test; and \textbf{explicitly}, by  using \MD{} (or similar) for proactive diversity.
}

\begin{table}[t!]
  \centering
\renewcommand{\arraystretch}{0.8}
    \caption{ Total and unique server detections among CC, libraries and FB for DPool \& QPool servers across all binaries}
    \setlength{\tabcolsep}{2.7pt}
  \setlength{\aboverulesep}{1.2pt}  
  \setlength{\belowrulesep}{1.2pt} 
  \footnotesize
  \begin{tabular}{c|c|c|c|c|c|c|c|c|c|c|c|c|c|c|c|c|c}
    \toprule
    & \multicolumn{6}{c|}{\textbf{Top-level CC}} & \multicolumn{5}{c|}{\textbf{Libraries}} & \multicolumn{6}{c}{\textbf{FB}} \\
    \cmidrule(lr){2-7} \cmidrule(lr){8-12} \cmidrule(lr){13-18}
    \textbf{Test} & 
    \rotatebox{90}{\textbf{Avx}} & 
    \rotatebox{90}{\textbf{Hasher}} & 
    \rotatebox{90}{\textbf{Pattern\_Gen}} & 
    \rotatebox{90}{\textbf{Malign\_Buff}} & 
    \rotatebox{90}{\textbf{Silkscreen}} & 
    \rotatebox{90}{\textbf{Utils}} & 
    \rotatebox{90}{\textbf{Zlib}} & 
    \rotatebox{90}{\textbf{OpenSSL}} & 
    \rotatebox{90}{\textbf{Abseil}} & 
    \rotatebox{90}{\textbf{Llvmlibc}} & 
    \rotatebox{90}
    {\textbf{Llvmlibc++}} & 
    \rotatebox{90}{\textbf{Swissmap}} & 
    \rotatebox{90}{\textbf{Proto}} & 
    \rotatebox{90}{\textbf{Libc}} & 
    \rotatebox{90}{\textbf{Tcmalloc}} & 
    \rotatebox{90}{\textbf{Hashing}} &
    \rotatebox{90}{\textbf{Stl-cord}}\\
    \midrule
    \textbf{Total} & 19 & 3 & 21 & 17 & 6 & 4 & 23 & 9 & 7 & 0 & 3 & 4 & 8 & 3 & 9 & 11 & 7 \\
    \midrule
    \textbf{Unique} & 15 & 0 & 7 & 3 & 3 & 0 & 4 & 1 & 0 & 0 & 0 & 1 & 3 & 0 & 3 & 1 & 3 \\
    \specialrule{\heavyrulewidth}{0pt}{0pt}
  \end{tabular}
  \label{tab:total-libraries}
\end{table}

\subsection{Reported 
Factors are Insufficient for Defect 
Detection and Reproducibility}
\label{Results4}

As established in \S\ref{Results3}, variation in execution context for failing instructions across programs explains the program-sensitivity of defect detection---why different programs executing the same failing instruction do not all detect the same defects. Lacking instruction-level visibility across diverse programs, prior work instead proposes \textit{instruction usage stress}---the dynamic frequency of a failing opcode in a test\mbox{---}as a predictor of detection, claiming %
detecting tests execute the defective opcode more frequently than non-detecting~ones~\mbox{\cite{alibaba-new,Wang:alibaba:sdc}}.

\begin{figure}[t!]
\centering
\includegraphics[trim={0 0 0 0.28cm},clip, width=1\linewidth]{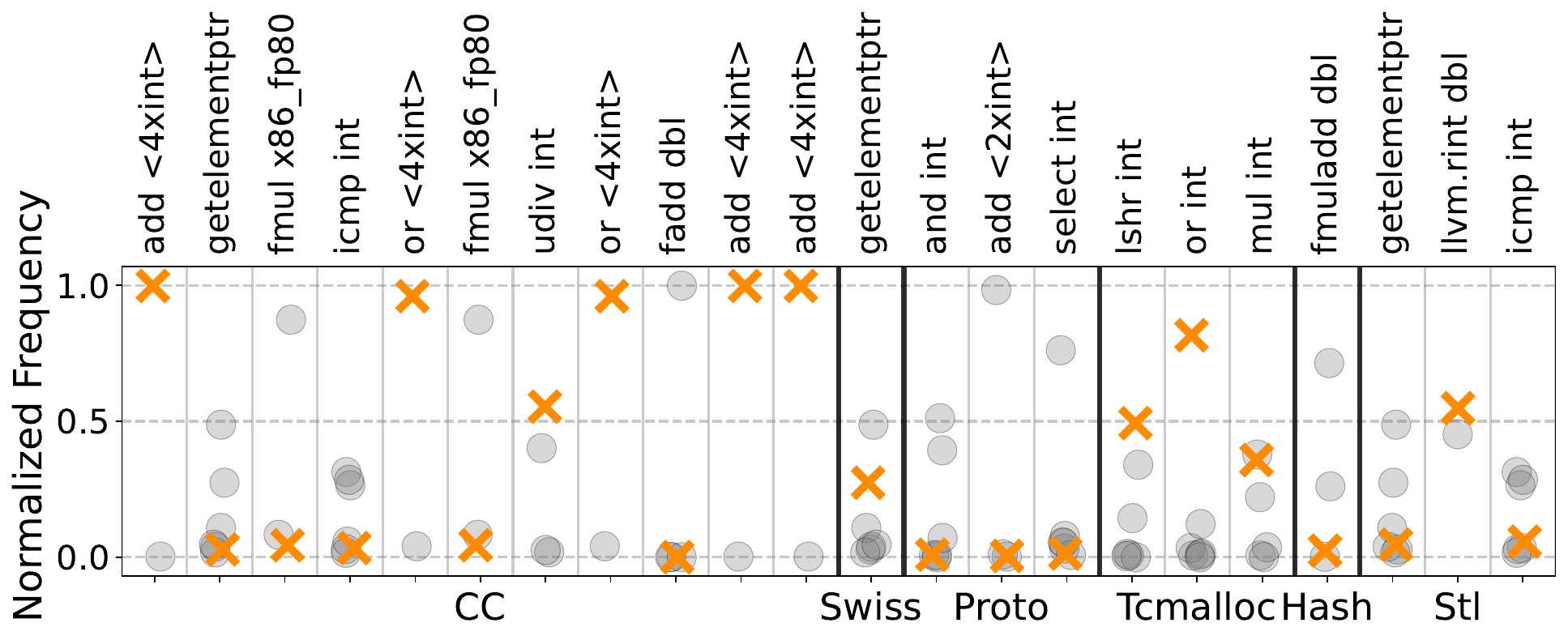}
\caption{Normalized 
execution frequency of failing
opcodes for servers (columns) uniquely detected by one %
program. %
Orange %
indicates~the~detecting program; gray
indicates non-detecting programs executing the same opcode.  
}
\label{fig:seq-dep}
\end{figure}

\looseness=-1
To investigate this claim, we focus %
on servers uniquely detected by a single test and for which at least one failing opcode also appears in at least one other test besides the detecting one. For each such server, if multiple failing opcodes are shared with the non-detecting tests, we select the most frequently failing one. 
Fig.~\ref{fig:seq-dep} shows the normalized execution frequency of this opcode across all tests that execute it. \obs{} In 13 of 22 cases (59\%), the detecting test is \textit{not} the one with the highest execution frequency of the failing opcode. 
This %
supports that execution context, %
not merely instruction usage stress, %
impacts error manifestation and therefore detection.
Notably, prior work also uses instruction usage stress for instruction localization, heuristically flagging as suspect those instructions that execute most frequently across failing testcases~\mbox{\cite{alibaba-new,Wang:alibaba:sdc}}~(Table~\ref{tab:related-work-hyperscalers}); Obs. 15 equally undermines this inference.

\begin{finding}
\textbf{Finding 7:}
\looseness=-2
The program-sensitivity of defect detection is not explained by %
the dynamic frequency of a %
failing opcode~(Obs.~15).
\end{finding}

Even for the same program running on the same server, not all runs %
detect errors: as Fig.~\ref{fig:heatmap} shows, detecting \CCithica{} tests have \texttt{EDR} less than 100\% for most servers. Two factors determine run-to-run error reproducibility: \textit{architectural}~(input-dependent) and \textit{non-architectural} (non-input-dependent) execution context.  
In the former case,  different runs use different randomly generated inputs, which influence error manifestation~(since instructions' inputs are part of their execution context) and detection (e.g., through logical masking), or even control flow (determining which instructions execute at all). %
In the latter case, microarchitectural  
and electrical state vary across runs of the same program on the same hardware~\cite{10xmitra}, 
causing errors to manifest in some runs but not others.

To assess the impact of non-architectural context on the reproducibility of detections across runs, we rerun \CCithica{} on three servers with high \texttt{EDR} (D3, D4, D7) with fixed inputs across runs, using \texttt{TTD}~(Table \ref{tab:metrics}) as a proxy for cross-run error reproducibility. 
Since the %
program inputs that resulted in  detections in earlier~experiments are not directly accessible, we use newly-generated random inputs. 
If the same input, and thus architectural context, repeatedly causes the same error to manifest, \texttt{TTD} should remain constant.

\looseness=-1
\obs{} We find that \texttt{TTD} distributions remain highly variable (e.g., for D3, ranging from seconds to nearly the full hour), demonstrating that %
non-architectural execution context contributes to error manifestation. Other servers did not exhibit any error detections in this experiment, precluding further analysis.%

\begin{figure}[t!]
\centering
\includegraphics[trim={0 0 0 0.0cm},clip, width=1.0\linewidth]{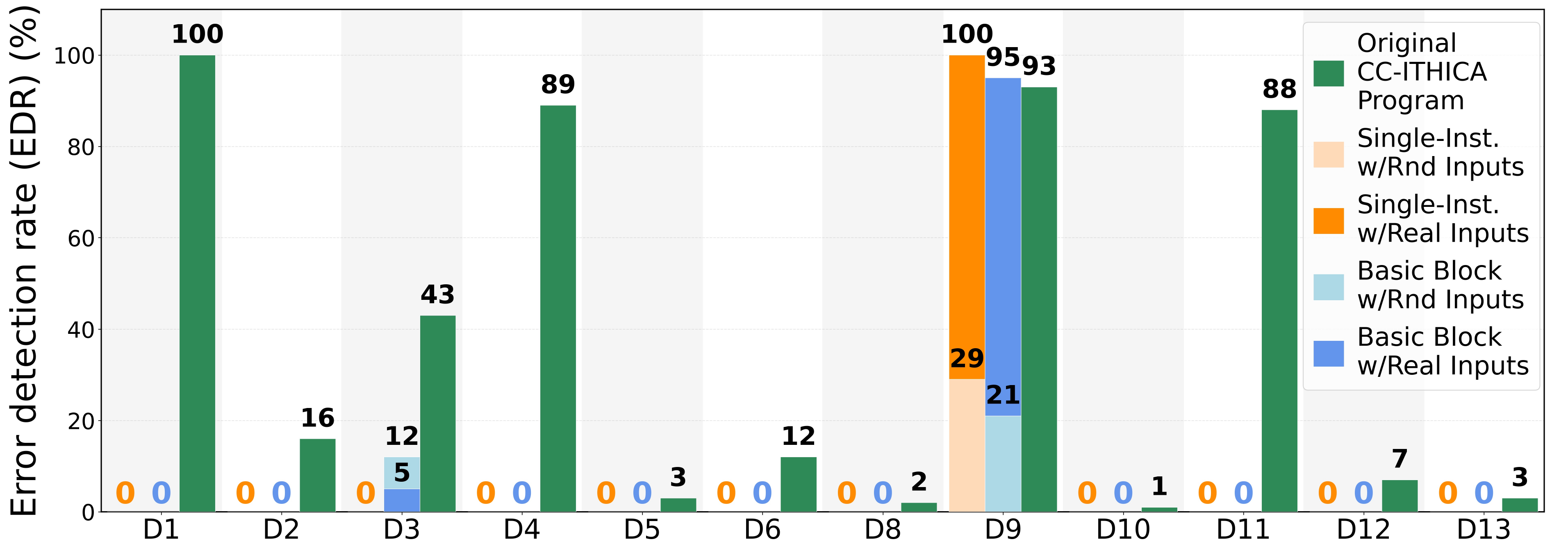}

\vspace{6pt}

  \setlength{\tabcolsep}{1.8pt}
  \renewcommand{\arraystretch}{0.9}
  \setlength{\aboverulesep}{0.7pt}
  \setlength{\belowrulesep}{0pt}
  \definecolor{darkorange}{RGB}{255,140,0}
  \definecolor{royalblue}{RGB}{65,105,225}
  \scalebox{0.78}{
  \begin{tabular}{@{}m{2.3cm}|m{0.7cm}|m{4.7cm}|m{1.45cm}|m{1.45cm}@{}}
    \toprule
    & \multicolumn{2}{c|}{\raisebox{-1pt}{\textbf{D3}}} & \multicolumn{2}{c}{\raisebox{-1pt}{\textbf{D9}}} \\    \cline{2-5}
    & \centering\raisebox{-1.5pt}{\textcolor{darkorange}{\textbf{Instr}}} & \centering\raisebox{-1.5pt}{\textcolor{royalblue}{\textbf{BB}}} & \centering\raisebox{-1.5pt}{\textcolor{darkorange}{\textbf{Instr}}} & \centering\arraybackslash\raisebox{-1.5pt}{\textcolor{royalblue}{\textbf{BB}}} \\
    \midrule
{Failing Opcode} & \multicolumn{1}{m{0.7cm}|}{\vspace{0.0pt}\centering None\par} & \multicolumn{1}{m{4.7cm}|}{\centering xor $<$4xint$>$ / trunc int / zext int \par} & \multicolumn{1}{m{1.45cm}|}{\centering fmul fp80\par} & \multicolumn{1}{m{1.45cm}}{\centering fmul fp80\par} \\
\hline
    Fail BB Length & \multicolumn{1}{c|}{---} & \multicolumn{1}{c|}{\makecell{ 17 / 13.5 / 25.0 }} & \multicolumn{1}{c|}{---} & \multicolumn{1}{c}{4.0} \\
    \hline
    \mbox{Pass BB Length} & \multicolumn{1}{c|}{---} & \multicolumn{1}{c|}{ \makecell{12.0 / 1.0 / 16.2 }} & \multicolumn{1}{c|}{---} & \multicolumn{1}{c}{---} \\
    \specialrule{\heavyrulewidth}{-1.3pt}{0pt}
  \end{tabular}
  }
  \caption{\textit{Top}: \texttt{EDR} of original \CCithica{}  program compared to single-instruction and single-basic-block reproducers (one %
  per failing instruction, averaged). Reproducers are evaluated under both %
  original failing inputs %
  (light bars) and random inputs (dark bars). \textit{Bottom}: For each reproducer, the failing opcodes and %
  lengths of failing vs. non-failing basic blocks.
  }
  \label{reproducibility-experiment}
  \label{tab:reproducer-details}
\end{figure}

\looseness=-1
To assess the length of instruction sequences required for reproducibility of defect-induced errors, 
we construct two additional types of tests per server:  
\textit{single-instruction tests} %
and \textit{basic-block tests}, each derived from original \CCithica{} programs %
by isolating a failing instruction or basic block. For each test, we evaluate both real failing inputs (from the original \CCithica{} runs) and newly-generated random inputs.
The basic-block tests range from 1 to 182 instructions in length, with a median of 15 and a mean of 32 instructions. 
 \A{} is used to generate tests for all servers except D2/D13, where  \M{}/\MD{} are used, %
which uniquely detected them.  Each test is run in a loop on each core  for \textbf{100 hours} (iterating through real or random inputs) and  the \texttt{EDR} results are reported in Fig.~\ref{reproducibility-experiment}, alongside the  \texttt{EDR} of the original \CCithica{} program from Fig.~\ref{fig:heatmap};  details for tests with detections are shown at the bottom. %

\obs{} Typically, neither single-instruction nor basic-block tests detect any errors, even
with real failing inputs.
There are two exceptions: 
D3~is not reproduced by single-instruction tests but is by basic-block tests, suggesting that modestly long sequences establish the necessary execution context for its detection. 
D9,~which fails on \textit{fmul x86\_fp80}, achieves perfect reproduction with the single-instruction tests with real inputs.

A potential explanation for D9 is reduced microarchitectural execution %
path non-determinism~\cite{RTL2MmPATH} for %
its failing LLVM IR  instruction.
Such non-determinism can arise from multiple sources: compilers may map an LLVM IR instruction to multiple assembly instructions; hardware may map an assembly instruction to multiple micro-ops; and hardware may assign micro-ops to one of multiple functional units. For D9, the first two sources are eliminated and the third is reduced: inspection of the compiled single-instruction LLVM IR reproducer reveals a single assembly instruction, \textit{fmul x86\_fp80}; \textit{fmul x86\_fp80} maps to a single micro-op per uops.info~\cite{uops-intel} and the server's microarchitecture documentation (undisclosed); and that micro-op can execute on one of two floating-point units (FPUs). 
The frequency
of \textit{both} original and validation instructions producing incorrect, inconsistent outputs 
suggests they often interact with the same defective hardware component. 
Since \ITHICA{} checks detect errors exclusively for \textit{fmul x86\_fp80} instructions across the full \CCithica{} test suite, this component is likely one of the FPUs.
However, %
hardware localization remains inherently opaque at the ISA level~(\S\ref{Results5}). Notably, even with this reduced microarchitectural execution path non-determinism and likely error persistence~(\S\ref{sec:qed-theory}),  
ITHICA still detects the error, making this case similar to our fault injection example~in~\S\ref{sec:rtl-fault-injection}.

This experiment, along with the \texttt{TTD} variability observed under fixed program inputs (Obs.~16), suggests that microarchitectural and electrical state, beyond architectural control~(\S\ref{execution-context-diversity}), contribute to defect-induced error manifestation.

\begin{finding}
\textbf{Finding 8:} 
\looseness=-1
Reproduction of defect-induced errors
generally cannot be done deterministically with
ISA-level~control~(Obs.~16,~17).
\end{finding}

Having just established that supplying the right inputs to the right opcodes is insufficient for reproduction, we further characterize the distribution of instruction-level input values that result in  error detections for each opcode-server combination for our full ITHICA programs, using \textbf{Input Breadth}  (Table~\ref{tab:metrics}). 
\obs{} Table~\ref{tab:pc_input_sensitivity} shows that, for most opcodes and servers, the number of unique failing inputs and Input Breadth are high, indicating that errors are largely not confined to %
a few \textit{culprit} input values.

\begin{finding}
\textbf{Finding 9:} 
For vulnerable instructions, selecting the
“perfect” input appears to be neither necessary nor
sufficient to induce a defect-induced error (Obs.~16--18).
\end{finding}

\implication{Implications for Chip-Users:}{%
\looseness=-2
Prior hyperscaler studies %
use~instruction usage stress as a predictor of defect detection and as a basis for instruction localization~\cite{Wang:alibaba:sdc,alibaba-new}; others~rely~on short %
tests for instruction localization~\cite{sevi-meta}. 
Our findings contradict both: instruction execution frequency does not reliably predict which tests detect a defective server, and therefore cannot reliably localize failing instructions, and far fewer errors are reproducible with short tests (quantified~for~our~experiments above) %
than with long ones~(Table~\ref{tab:related-work-hyperscalers}, Conclusion~1).
}

\begin{figure}[t!]
\centering
\includegraphics[trim={0cm 0.0cm 0.0cm 0.0cm},clip, width=1\linewidth]{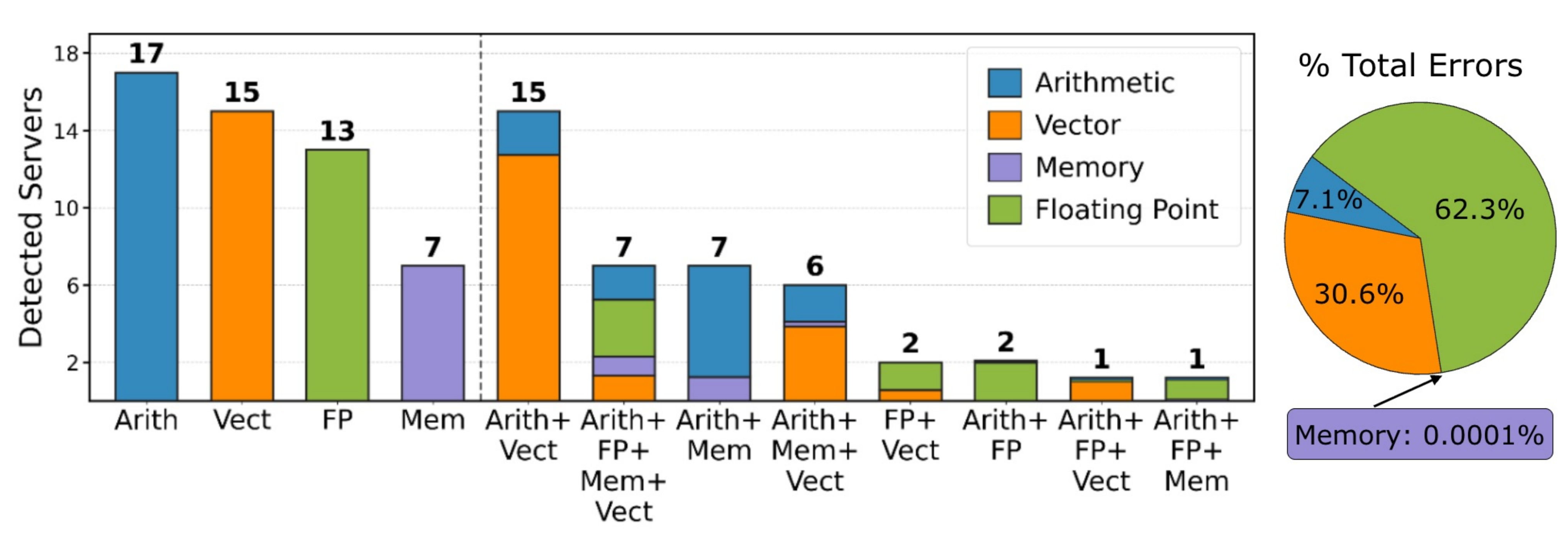}
\caption{Combinations of failing instruction types across ITHICA tests. Color segments reflect average  per-server error breakdown, %
along with overall error breakdown (pie chart).}
\label{fig:combinations} 
\end{figure}

\subsection{Hardware Localization from the ISA Level is Opaque} 
\label{Results5}

Fig.~\ref{fig:instructions} shows that some DPool servers exhibit errors concentrated in a single instruction type (e.g., D10), while others show errors across types (e.g., D1, D12). 
Fig.~\ref{fig:combinations} extends this analysis by
examining the distribution of errors across instruction types  
for 93 of all 100 ITHICA-detected servers (89 by \CCithica{} and 11 unique to \FBithica{}); seven are excluded from this analysis due to a crash that prevented the logging of the failing instruction from completing (\textit{no-instr},~\S\ref{Results1}). %
For each combination of instruction types, the figure shows the number of servers exhibiting errors for that combination and the average per-server breakdown of number of errors by instruction type. %
The %
pie chart aggregates this breakdown across all 93 servers.

\obs{} Among all 93 servers analyzed, 44\%  exhibit errors~spanning multiple instruction types, %
rather than a single type, consistent with prior work observing multi-family test failures~\cite{pindrop,alibaba-new,Wang:alibaba:sdc}; we newly observe this phenomenon at the instruction level and quantify the rate disparity between types.
\obs{} Despite accounting for a large share of detected servers, arithmetic and memory instructions contribute only a small fraction of errors: 92.9\%  originate from floating-point and vector instructions.
In particular, floating-point instructions exhibit errors six orders of magnitude more frequently than memory instructions, on average. 
These observations suggest that %
floating-point and vector instructions may be easier to reproduce or, equivalently, have reduced microarchitectural execution path non-determinism~(\S\ref{Results4}).

\looseness=-1
ITHICA's %
ability to detect errors across multiple instruction types within the same test illustrates why drawing hardware localization conclusions from 
tests that check only a narrow set of instructions
is problematic.
\obs{} For example, the only baseline test that detects D1 is the %
\texttt{Avx} subtest of CC, which is a vector-specialized test. From this test alone, one may blame a vector-specific hardware unit.  Yet ITHICA's instruction-level checks within the same test reveal %
non-vector instructions are also affected, %
suggesting a different cause. Overall, in our experiments, vector instructions
exhibit errors on 46 servers,~31~of which feature errors in other instruction types as well. This challenges prior work's hardware localization conclusions   that attribute errors occurring in  vector instructions to 
vector units, using vector-specific tests alone~\cite{sevi-meta}.

\begin{finding}
\textbf{Finding 10:}
 The same defect often causes errors across %
 instruction types, at markedly different rates (Obs.~19--21). %
\end{finding}

\begin{table}[t!]
\caption{Comparison with related hyperscaler fleet studies 
}
\setlength{\tabcolsep}{1pt}
\renewcommand{\arraystretch}{0.5}
\setlength{\aboverulesep}{0.5pt}
\setlength{\belowrulesep}{0.5pt}
\scalebox{0.792}{%
\begin{tabular}{@{}c | l | c | c | c@{}}
\toprule
&
& \makecell{\textbf{Alibaba}~\cite{Wang:alibaba:sdc,alibaba-new}}
& \makecell{\textbf{SEVI}~\cite{sevi-meta}}
& \makecell{\textbf{ITHICA}~\emph{(this work)}} \\
\midrule
\multirow{2}{*}[\dimexpr+5.8pt]{\rotatebox[origin=c]{90}{\textit{Evaluation}}}
& \makecell[l]{Suspect-Defective\\Pool Size}          & Unclear       & $>$2,500      & $>$3,000 \\
\cmidrule(l){2-5}
& \makecell[l]{Servers Detected\\\& Analyzed}  & 27--30            & 18            %
& \textbf{100}\\
\midrule[0.8pt]
\multirow{4}{*}[\dimexpr-2pt]{\rotatebox[origin=c]{90}{\textit{Testing Method}}}
& \makecell[l]{Error\\Assumption}           & \makecell{Consistent\\(implicit)}  & \makecell{Consistent\\(explicit)} & \makecell{Inconsistent\\(validated on real HW, \S\ref{Results1})} \\
\cmidrule(l){2-5}
& \makecell[l]{Inst. Types Checked}       & \makecell{--- (not inst.-level)} & Vector only  & \makecell{All  (Arith, FP, Vec, Mem, CF)} \\
\cmidrule(l){2-5}
& \makecell[l]{Error Check\\Technique}           & \makecell{Final output w/\\golden value} & \makecell{Inst.-level w/ \\golden value} & \makecell{Intra-thread\\inst.-level} \\
\cmidrule(l){2-5}
& \makecell[l]{Instruction\\Localization}        & %
\makecell{Post-detection via\\inst. usage stress} & 
\makecell{Concurrent w/\\detection} & \makecell{Concurrent w/\\detection} \\
\midrule[0.8pt]
\multirow{4}{*}[\dimexpr-8pt]{\rotatebox[origin=c]{90}{\textit{Conclusions}}}
& \makecell[l]{1. Factors\\Determining\\Detection}             & \makecell{Inst.  usage stress,\\Temperature}  & \makecell{Opcode, Input,\\Temperature} & %
\makecell{%
Sequence-driven  execution\\context
(\S\ref{Results3}); Inst. usage \\  stress \& opcodes/inputs\\ alone are insufficient (\S\ref{Results4})} \\
\cmidrule(l){2-5}
& \makecell[l]{2. Vulnerable\\HW Unit}        & \makecell{ALU, FPU, Vec.Unit,\\Cache, TrxMem} & \makecell{Vector\\multiplier} & \makecell{Shows prior work conclusions \\ are unreliable %
(\S\ref{Results5})} \\
\cmidrule(l){2-5}
& \makecell[l]{3. Solution\\Proposed}         & \makecell{Can focus on\\vulnerable features} & \makecell{ABFT for\\matmul kernels} & \makecell{Diverse programs as tests (\S\ref{Results3}); \\ Inst. localization concurrently\\ with detection (\S\ref{Results4})} \\
\bottomrule
\end{tabular}%
}
\label{tab:related-work-hyperscalers}
\end{table}

More
generally, while ITHICA’s instruction-level visibility enables \textit{scrutinizing} hardware localization conclusions, it cannot definitively \textit{establish} them. This is a limitation of 
all ISA-level testing approaches~\cite{google:silifuzz,cpu-check,alibaba-new,Wang:alibaba:sdc,sevi-meta,pindrop,harpocrates,intel:opendcdiag}, including ITHICA. A defect may reside \textit{anywhere} along an instruction's microarchitectural  execution path or in physically adjacent transistors serving unrelated hardware components, none of which is visible at the ISA level. Moreover,
the compiler's mapping of LLVM IR to assembly, and the hardware's mapping of assembly to micro-ops and micro-ops to functional units, introduce microarchitectural execution path non-determinism %
inherent to 
 ISA-level approaches~(\S\ref{Results4}).

\implication{Implications for Chip-Users:}{ 
Hardware localization 
is inherently opaque from ISA-level observations. 
Although ITHICA's instruction-level checking across diverse instructions and programs does not overcome this opacity, it surfaces 
the full range of LLVM instructions affected on each server, from which prior hardware localization conclusions~\cite{sevi-meta,Wang:alibaba:sdc,alibaba-new} (Table~\ref{tab:related-work-hyperscalers}, Conclusion 2) can be scrutinized and, in some cases, discredited.  
Incorrect hardware localization  conclusions  risk guiding future software and hardware mitigations toward the wrong or insufficient components~(Table~\ref{tab:related-work-hyperscalers}, Conclusion~3).}

\section{Comparison with SiliFuzz}
\label{fuzzing}
\begin{figure}[t!]
    \centering
    \begin{minipage}[c]{0.52\linewidth}
        \includegraphics[width=\linewidth]{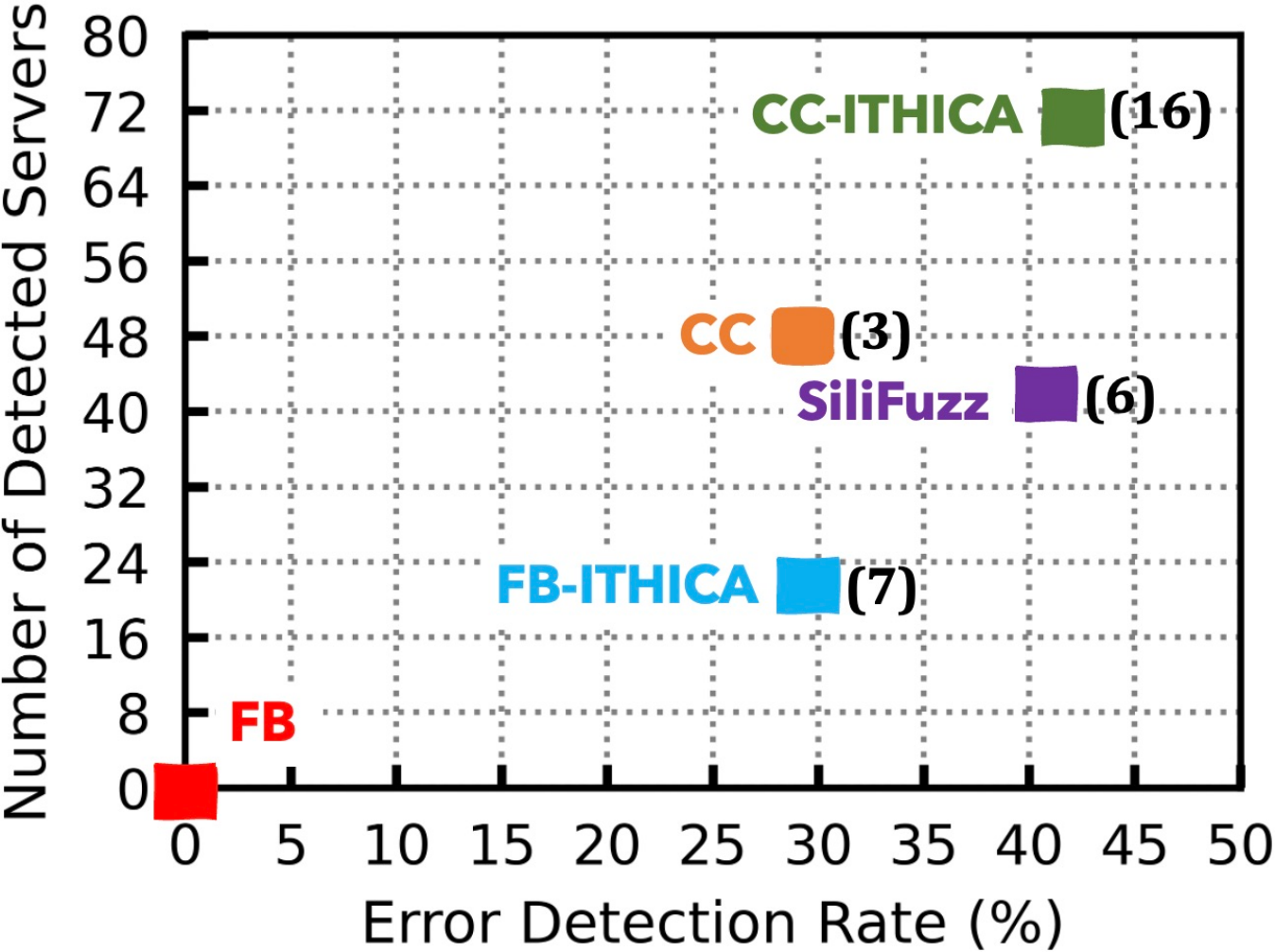}
    \end{minipage}%
    \hfill%
    \begin{minipage}[c]{0.44\linewidth}
        \caption{
        Comparison of SiliFuzz, \CC{} and ITHICA
         tests
         across all commonly tested servers. Unique server detections for each test (missed by all other tests)
        are shown in parentheses.}
        \label{figure:comparison-with-silifuzz}
    \end{minipage}
\end{figure}

Fig.~\ref{figure:comparison-with-silifuzz} compares \CC{}, \CCithica{}, and \FBithica{} with SiliFuzz~\cite{silifuzz-github,google:silifuzz}, the  
only fuzzing tool for defects that does not rely on below-ISA specifications~(\S\ref{sec:related-work}). The tests are compared  
by number of total and unique detected servers and average \texttt{EDR}. %
\texttt{TTD} and \texttt{EF} statistics are not available for SiliFuzz and are therefore omitted.  Similar to~\S\ref{sec:experiment2}, we exclude servers with less than 20 execution hours per test.  
For each remaining server, we use an equal number of one-hour executions per test, set to the minimum completed by any single test on that server.
SiliFuzz detects significantly fewer servers than \CCithica{} (42~vs.~71). For the servers it  detects, its average \texttt{EDR} is smaller than but comparable to \CCithica{} (40.8\%~vs.~42.1\%). Notably, \FBithica{} detects seven servers missed by all other tests.

\section{Limitations}
\label{limitations}
This section discusses limitations of our fleet study in general and ITHICA tests in particular.

\myparagraph{Server Pool Bias} The servers evaluated in our %
study represent a biased sample of \Company{}'s overall fleet. 
Our results in \S\ref{results}--\S\ref{fuzzing} are derived from the QPool and DPool, which comprise suspect- and confirmed-defective servers, respectively~(\S\ref{sub:settup}). 
Because these out-of-production pools are populated using non-ITHICA screening mechanisms, they inherently over-represent servers with defects that those existing tests can flag.
This selection bias likely explains why ITHICA's \texttt{Arith} checks dominate our experimental results (\S\ref{Results1}), given that many standard screening tests explicitly stress arithmetic operations~\cite{cpu-check,google:silifuzz}. 
We note, however, that this bias is standard across the literature: prior hyperscaler fleet studies~\cite{pindrop, sevi-meta, Wang:alibaba:sdc, alibaba-new, dixit:sdc-detecting} evaluate similarly-populated pools, suggesting our study examines a comparable server~population.

\myparagraph{Missing Defect-Induced Errors} 
ITHICA tests can miss defect-induced errors---e.g., those on the six servers in Fig.~\ref{fig:qpool-new} that \texttt{ITHICA} checks miss but \Native{} checks detect---because of instrumentation gaps or because they manifest consistently.

\textit{Instrumentation Gaps:} 
Even at maximum instrumentation~(block size of one), ITHICA does not instrument, and thus does not check, every LLVM IR instruction executed in the program it transforms into a test. For example: (1)~instructions with side-effects (i.e., that are not safe to duplicate), including~atomic and volatile memory instructions~(\S\ref{sec:llvm-implementation}), %
(2)~memory instructions in thread-unsafe code~(\S\ref{sec:llvm-implementation}), (3)~uninstrumented parts of %
\texttt{Libc}/\texttt{Libc++}~(\S\ref{sub:software-setup}), %
and (4) kernel code executed as part of system calls. Gaps~(1)--(3) apply only to \ITHICA{} checks, while~(4) is shared between \ITHICA{} and \Native{} checks.  
Despite~(2), ITHICA still instruments most (non-memory) opcodes in thread-unsafe programs, which are difficult to check with prior error checking techniques~(\S\ref{sec:intro}). %

\textit{Consistent Errors:}
By design, \texttt{ITHICA} checks detect inconsistent errors only, not consistent ones; the \Native{} checks in \CC{}/\CCithica{} can in principle detect both. ITHICA's focus stems from our %
insight that consistent errors are theoretically easier to detect, since they manifest from every execution context, whereas inconsistent errors manifest from only select ones~(\S\ref{sec:qed-theory}). We show that it pays off empirically: \texttt{ITHICA} checks detect 39\% more defective servers than \Native{} checks within the same test programs, and \texttt{ITHICA} checks in non-test workloads detect additional unique defective servers.

\myparagraph{Heuristic Coverage} 
\looseness=-1
Finally, ITHICA tests can fail to induce an error on a defective server
due to ITHICA
neither measuring nor controlling \textit{hardware coverage}---the %
fraction of distinct microarchitectural and electrical state contexts~(\S\ref{key-insight}) and thus potential fault sites. %
As a result, its tests may never exercise a server's defect site in a context where the defect manifests. 
Today, no metric is known to accurately measure hardware coverage of functional tests for defects: %
existing  
metrics---structural coverage~\cite{Wadsack:stuck-open,Waicukauski:transition-fault,
Smith:path-delay-fault,Heragu:segment-delay-fault,Storey:bridging-fault,
Ma:n-detect,Hapke:cell-aware-fault,McCluskey:gate-exhaustive-fault}, ACE-based~\cite{avf,harpocrates}, or delay-fault~\cite{delay-avf} vulnerability---all
score tests against fault models %
that have not been shown to
accurately represent
defects~\cite{McCluskey:stuck-fault,Wei:pepr,10xmitra,intel-screening-manuf-defects}.

Our evaluation therefore reports detection outcomes~(Table~\ref{tab:metrics}) instead, as do
all prior fleet studies~\cite{Dixit:2021:sdc-at-scale,dixit:sdc-detecting,
Wang:alibaba:sdc,alibaba-new,pindrop,sevi-meta}. %
Only two prior techniques for defects define and optimize a coverage metric, but neither is shown to predict detection outcomes on real defective hardware.
SiliFuzz~\cite{google:silifuzz} maximizes control-flow coverage of a software CPU
emulator as a proxy for hardware coverage, and acknowledges the two do not necessarily  correspond;
empirically, ITHICA detects~69\% more servers than SiliFuzz~(\S\ref{fuzzing}). %
Harpocrates~\cite{harpocrates} defines
its own coverage metric for defects, but evaluates its correlation  with detection outcomes (i.e., errors detected) using stuck-at
fault injection rather than real defective hardware. %
Therefore, like other functional tests, ITHICA tests
 cannot \textit{systematically} improve coverage of %
 defect fault sites;
instead, they \textit{heuristically} improve it through the test-diversity levers
of~\S\ref{Results3}, validated on real defective~hardware.

\section{Related Work}
\label{sec:related-work}

Prior works most relevant to ITHICA fall into four categories: %
soft error detection, post-silicon validation, %
fuzzing, %
and 
hyperscaler fleet studies %
of the CPU SDC problem.
Table~\ref{tab:related_work} compares \approach{}
to existing functional test suites~\cite{cpu-check,intel:opendcdiag} and test generation techniques~\cite{google:silifuzz,harpocrates} for defects.

\myparagraph{
Soft Error Detection} %
\label{soft-error-techniques}
EDDI~\cite{EDDI} is the first to propose intra-thread instruction checks via instruction duplication and output comparison.  %
CFCSS~\cite{CFCSS} is the first to introduce intra-thread checks %
for branches; it does so by validating  
runtime control-flow against compile-time control-flow graphs. %
SWIFT~\cite{Swift} reduces EDDI's performance overhead by %
 omitting memory instruction  protection (relying on ECC~\cite{ecc})  
and by 
reusing idle processor resources for duplicate computation,  
and improves CFCSS's control-flow coverage.  %
Several variations %
of the above techniques propose further %
performance or coverage 
improvements, %
either purely in software~\cite{fast-low-level24,nZDC,Shoestring,Multiple_inputs,ACFC,IPAS,lu2014sdctune,esoftcheck-yu2009}  or with architectural support~\cite{CRAFT,implicit-signature-checking-hardware}. Analogous software-only~\cite{ArmorAllGPU,optimizing-sc-gpu,approxdup} and architecture-supported~\cite{swapcodes,optimizing-sc-gpu} %
techniques have been proposed %
for GPUs. %

Unlike prior soft error detection work, whose goal is application protection during in-production execution, ITHICA's goal is \textit{eventual} defect detection during out-of-production testing. This distinction affects how error checks are inserted---for which instructions (sampling in space) and how frequently~(sampling in time)---and the relevance of non-check program-diversity levers~(\S\ref{Results3}), as described below.  %

\textit{Error Check Insertion:}
Sampling in time is applicable to ITHICA but not soft error detection. The former requires eventual defect detection, so it can skip checks (e.g., by increasing block size,~\S\ref{Results2}); the latter must check every instruction that can affect a program's output to guarantee fault tolerance.
Sampling in space is applicable to both, but relevant mechanisms for achieving it differ. %
For example, soft error detection can
exploit logical masking to prune error checks for instructions
deemed incapable---either by static program analysis~\cite{Shoestring,esoftcheck-yu2009} or by
characterization on known inputs~\cite{IPAS, lu2014sdctune}---of causing an
error at a program’s output. Such pruning is inapplicable to ITHICA, since logical
masking harms its defect detection capabilities~(\S\ref{sec:llvm-implementation},~\S\ref{Results2}).

\textit{Diversity Levers:}
ITHICA's non-check levers~(\S\ref{Results3}) diversify how a program exercises the hardware, %
heuristically improving defect
detection. These levers are irrelevant to soft error protection, which addresses transient rather than permanent faults and seeks to prevent such errors~altogether.

\myparagraph{Post-Silicon Validation}  
QED~\cite{Hong:2010:qed,lin2013overcoming,Lin2014Effective,Singh:2017:eqed} adapts its intra-thread instruction checks from  soft error detection
to detect logic and electrical design bugs during post-silicon validation. Thus, QED is intended to be run on relatively few (hundreds~\cite{dixit:sdc-detecting}) silicon %
samples before mass production. Per \S\ref{sec:ithica-passes}, ITHICA further adapts these checks for defect detection and demonstrates their effectiveness in this new setting %
\textit{for the first time}. 
In contrast to QED, ITHICA is intended to be used on millions of deployed servers during in-datacenter testing. These differences in fault scopes and deployment scenarios between QED and ITHICA lead them to adopt different program transformations and make different implementation~choices. %

\textit{Program transformations:}
Driven by our insight that execution context diversity is important for defect detection~(\S\ref{execution-context-diversity}), ITHICA uses several levers to increase it, 
including the novel \MD{} transformation, which detects unique servers in our experiments~(\S\ref{Results3}).  %
It is possible for \MD{}~to~benefit post-silicon validation, though it has not been evaluated for that purpose.
Additionally, %
QED implementations~of control-flow checking transformations
can~leverage known~test inputs---a reasonable assumption for post-silicon validation.  ITHICA’s \B{}
cannot, and does not, as it operates on arbitrary~programs.%

\textit{Implementation choices:}
QED leaves library code uninstrumented, due to its source-code-level implementation.    %
ITHICA’s LLVM IR passes, by comparison, can instrument~any code compiled through LLVM, including
libraries---where ITHICA detects %
unique servers in our experiments~(\S\ref{Results3}).

\begin{table}[t!]
  \centering
  \setlength\tabcolsep{1.1pt}
  \renewcommand{\arraystretch}{1.0}
  \setlength{\aboverulesep}{2pt}  
  \setlength{\belowrulesep}{1.4pt}
  \caption{
Tests and test generation techniques for defects 
}%
\footnotesize %
  \begin{tabularx}
  {0.48582\textwidth}{
    |c %
    |c|c|c|c %
    |c|c %
    |c| %
    }
    \hline
    \multirow{2}{*}{\raisebox{-3mm}{\textbf{Technique}}} &
    \multicolumn{4}{c|}{\textbf{Check Type}} &
    \multicolumn{2}{c|}{\textbf{Evaluation}} &
    \multirow{2}{*}{\raisebox{-3.1mm}{\makecell{\textbf{Spec}}}} \\
    \cline{2-7}
    &
    \makecell{\textbf{Golden}\\[-2pt]\textbf{Value}} & 
    \makecell{\textbf{Cross-}\\[-2pt]\textbf{Core}} & 
    \makecell{\textbf{Invert.}\\[-2pt]\textbf{Comp.}} &
    \makecell{\textbf{Intra-thr.}\\[-2pt]\textbf{Instruct.}} &
    \makecell{\textbf{Fault}\\[-2pt]\textbf{Inject.}} & \makecell{\textbf{Real}\\[-2pt]\textbf{HW}} & \\
    \thickhline
    {cpu-check}~\cite{cpu-check} &
    {\whitexmark} & {\bluecmark} & {\bluecmark} & {\whitexmark} &
    \textbf{--} & \textbf{--} &
    C/C++ \\
    \hline
    {OpenDCDiag}~\cite{intel:opendcdiag} &
    {\bluecmark} & {\bluecmark} & {\bluecmark} & {\whitexmark} &
    {\whitexmark} & {\bluecmark} &
    C/C++ \\
    \thickhline
 {SiliFuzz}~\cite{google:silifuzz} &
    {\bluecmark} & {\bluecmark} & {\whitexmark} & {\whitexmark} &
    \textbf{--} & \textbf{--} &
    ASM \\
    \hline
{Harpocrates}~\cite{harpocrates} &
    {\bluecmark} & {\bluecmark} & {\whitexmark} & {\whitexmark} &
    {\bluecmark} & {\whitexmark} &
    Uarch \\
    \thickhline
    {\raisebox{-0.6pt}{\textbf{ITHICA} \textit{(this work)}}} &
    {\whitexmark} & {\whitexmark} & {\whitexmark} & {\bluecmark} &
    {\whitexmark} & {\bluecmark} &
    {\raisebox{-0.6pt} {LLVM}} \\
    \thickhline
  \end{tabularx}

\label{tab:related_work}
\end{table}

\myparagraph{Fuzzing} A large number of fuzzing works exist, targeting  defects, design bugs, or security bugs~\cite{pathfuzz,difuzzrtl,thehuzz,harpocrates,fuzzing-hardware-like-software,syzkaller,google:silifuzz}; several of them rely on below-ISA specifications~\cite{pathfuzz,difuzzrtl,thehuzz,harpocrates}. %
Closest in motivation to ITHICA is PathFuzz~\cite{pathfuzz}, which uses real-program checkpoints as fuzzer seeds for pre-silicon validation, but, unlike ITHICA, requires deterministic seeds, restricting usable programs. %
More
broadly, ITHICA checks could 
be applied to fuzzer-generated sequences to replace the golden outputs they require~\cite{pathfuzz,harpocrates,thehuzz,difuzzrtl,google:silifuzz}.

\myparagraph{Hyperscaler Fleet Studies} Like ITHICA, the Alibaba studies~\cite{Wang:alibaba:sdc,alibaba-new} and SEVI~\cite{sevi-meta} %
evaluate their techniques using pools of suspect-defective servers. %
Table~\ref{tab:related-work-hyperscalers} compares their testing methods and findings with ITHICA's. %
Other works %
quantify the prevalence and report detection trends of
hardware errors across large CPU fleets~\cite{Dixit:2021:sdc-at-scale,dixit:sdc-detecting,Hochschild:2021:cores-dont-count,10xmitra,pindrop,hardware-sentinel-asplos25} or study SDC vulnerability through fault injection~\cite{veritas,giz-perspectives,giz-measure2024,saboteurs}. Among them, PinDrop~\cite{pindrop} advocates for continuous testing with a variety of complex tests at scale---a philosophy that aligns with ITHICA's goal of enabling arbitrary programs to be used as tests. Mitra et al.~\cite{10xmitra} finds that a significant fraction of defects %
are %
detected by production workloads in the field, motivating the instrumentation of such workloads with ITHICA.

\section{Conclusion}
We present \approach{}, an approach for  automatically generating functional tests from arbitrary programs %
to detect defective CPUs. \approach{} exploits the insight that the most pernicious defects manifest as inconsistent errors---validated here for the first time on real hardware---to build intra-thread, instruction-level checks that outperform baseline industry checks and enable novel findings on defect behavior 
that challenge conclusions of prior hyperscaler studies.

\section*{Acknowledgment}
This work was supported in part by the U.S. National Science Foundation under award 2321489 and a Sloan Research Fellowship. We also gratefully acknowledge gifts from Google, Meta, and the Open Compute Project. Finally, we thank Peter Hochschild, David Culler, Parthasarathy Ranganathan, Liqun Cheng, Martin Dixon, Konstantin Shtoyk, and Rama Govindaraju for their continued~support.

\bibliographystyle{IEEEtran}
\bibliography{refs}

\clearpage

\appendix
\section{Artifact Appendix}
\emergencystretch=1.5em
\ifdefined\setlist\setlist{itemsep=1pt,topsep=2pt,parsep=0pt}\fi

\makeatletter
\newcommand{\code}[1]{{\fontfamily{zi4}\selectfont\relsize{-0.5}#1}}
\newcommand{\aecodesetup}{%
  \topsep=2pt\partopsep=0pt\small
\renewcommand{\verbatim@font}{\normalfont\fontfamily{zi4}\selectfont}%
  \@noligs
}
\makeatother

\subsection{Abstract}

This artifact includes %
the ITHICA LLVM IR passes (\A{}, \M{}, \MD{}, \B{} and their combinations) used to transform real programs into tests for hardware defects (first contribution in~\S\ref{sec:intro}). It demonstrates that ITHICA passes successfully instrument the two programs used in the paper: cpu-check~(\CC{}) and Fleetbench~(\FB{}). For ease of use, the Zenodo artifact provides a pinned Docker build environment per program (kept separate due to disjoint build requirements), containing all dependencies needed to compile both programs (with and without ITHICA) from source. The artifact  reproduces the runtime  performance overhead and binary size increase that ITHICA transformations impose relative to the uninstrumented \CC{} baseline,  equivalent to Table~\ref{tab:overheads}. Pre-built binaries  for all configurations of Table~\ref{tab:overheads} are provided to  reproduce the  measurements directly, even without recompilation. The remaining results~(\S\ref{results}--\S\ref{fuzzing}) require access to proprietary defective hardware owned by Google, and therefore cannot be reproduced.  %

\subsection{Artifact check-list (meta-information)}

{\small
\begin{itemize}
  \item {\bf Transformations:}  Code for ITHICA's LLVM IR compiler passes, \A{}, \M{}, \MD{}, and \B{}, and their combinations \AM{}, \AMD{} and \AMDB{}.

  \item {\bf Program:} The input programs compiled with ITHICA: %
  cpu-check and  Fleetbench. Both are open-source and shipped with the artifact. %

  \item {\bf Compilation:} No dependencies beyond Docker. Clang/LLVM/lld 14.0.6 %
  and Bazel 7.3.1 are used within the two Docker images. %

  \item {\bf Binary:} Pre-built, statically linked x86-64 binaries available.

  \item {\bf Hardware:} 
  Required: x86-64 Linux machine  %
  with at least 16 GB of RAM %
  and 40 GB of free disk space on the Docker filesystem.
  Preferred: 32 cores or more and hyperthreading enabled. %
  (Reference host: Intel Xeon Gold 6226R; 2$\times$16 cores, 2 threads/core, 500 GB RAM).

   \item {\bf Runtime Environment:} Docker and Bash. Network access.

  \item {\bf Metrics:} Runtime performance overhead and binary size increase relative to the %
  baseline. %

  \item {\bf Experiments:} %
  Two experiments (\S\ref{sec:artifact-eval}): E1 reproduces the results of Table~\ref{tab:overheads}---it measures per-ITHICA-pass overhead and binary-size increase on \CC{}, including %
  parameter sweeps for \A{}; E2 measures the same metrics on \FB{} with and without \A{} (not in the main paper). Separately, this artifact allows rebuilding both programs from source. %

  \item {\bf How much disk space required (approximately)?:} 40 GB. 
  
  \item {\bf How much time is needed to prepare workflow
        (approximately)?:} 
        10 minutes for Docker environment setup. %
        To recompile all binaries from source: additional 8~hours.

  \item {\bf How much time is needed to complete experiments
        (approximately)?:} 
        65 minutes total to execute all binaries. %

  \item {\bf Publicly available?:}   \href{https://github.com/ioannavav/ithica}{https://github.com/ioannavav/ithica} 
  \item {\bf Code licenses (if publicly available)?:} 
        MIT License. 
        
  \item {\bf Workflow automation framework used?:} Docker. %
        
  \item {\bf Archived (DOI)?:} %
\href{https://doi.org/10.5281/zenodo.21554153}{https://doi.org/10.5281/zenodo.21554153}  
\end{itemize}
}
\subsection{Description}

\subsubsection{How to access}\label{sec:artifact-access}
The artifact is archived on \href{https://doi.org/10.5281/zenodo.21554153}{https://doi.org/10.5281/zenodo.21554153} as a complete, ready-to-run copy (pre-built binaries and toolchain image included).   The source code is also on  \href{https://github.com/ioannavav/ithica}{https://github.com/ioannavav/ithica}. 

The artifact contains two directories, \code{cpu-check-ae/} and \code{fleetbench-ae/}, one per program. Each holds the ITHICA pass sources, the pinned program source, the pinned toolchain, pre-built binaries (to avoid heavy recompilation of some programs if needed), and three driver scripts: \code{setup.sh} provisions the toolchain, \code{build.sh} compiles the program with a chosen ITHICA pass, and \code{measure.sh} runs the (resulting or pre-built) binaries and reports the metrics. %

\subsubsection{Hardware dependencies}\label{sec:artifact-hw}
x86-64 Linux machine  %
 with 40 GB of free disk space and $16$\,GB RAM. %
To reproduce the reported results closely, %
a host with at least 32 physical / 64 logical CPUs is recommended.
Reference host: Intel Xeon Gold 6226R: 2 sockets $\times$ 16 cores \emph{per socket} $=$ 32 physical cores, and 2 threads per core $=$ 64 logical CPUs; 500\,GB RAM.
The results obtained are expected to vary depending on the underlying hardware~(\S\ref{sec:artifact-notes}).

\subsubsection{Software dependencies}\label{sec:artifact-sw}
  Only Docker (the cpu-check toolchain is exported from a pinned GNU~Guix environment via  \texttt{guix} \texttt{pack} and the Fleetbench image is built from a Dockerfile). 
  Network access is required while building either image. %
  On hosts where Docker's bridge network cannot reach GitHub, add \code{-{}-network=host} to \code{docker}~\code{build}.

\subsection{Installation}

Use the Zenodo copy for the installation steps below and results reproduction using pre-built binaries.   
The build steps in \S\ref{sec:artifact-eval} for all programs use these images:
\begin{itemize}
\item \emph{For cpu-check:}
{\aecodesetup\begin{verbatim}
cd cpu-check-ae
scripts/setup.sh   # docker load (shipped image)
\end{verbatim}
}
\item \emph{For Fleetbench:}
{\aecodesetup\begin{verbatim}
cd fleetbench-ae
scripts/setup.sh   # docker build (Dockerfile)
\end{verbatim}
}
\end{itemize}

\subsection{Experiment workflow}\label{sec:artifact-workflow}
The evaluation first compiles the two programs with  ITHICA;   \code{build.sh} builds the target pass and compiles the program in one step; \code{build\_pass.sh} builds a pass separately. 
Two script-driven experiments then measure the
resulting binaries: for cpu-check, E1 measures the per-pass overhead
and binary size increase compared to the uninstrumented baseline (main result, reproducing Table~\ref{tab:overheads}), and E2 measures the same two
metrics for Fleetbench with and without \A{} (with block size and interleaving of 1).
A toy program (\code{toy\_overhead/toy\_program.cpp}) ships in
the cpu-check image so users can quickly instrument and run it
(\code{cpu-check-ae/scripts/toy\_overhead.sh}).

\subsection{Evaluation and expected results}\label{sec:artifact-eval}

\begin{table}[t!]
  \centering
\renewcommand{\arraystretch}{0.8}
  \setlength{\aboverulesep}{1pt}  
  \setlength{\belowrulesep}{1pt}    
  \setlength{\tabcolsep}{2pt}
  \footnotesize
    \caption{
  Results equivalent to %
  Table~\ref{tab:overheads}, here measured on a non-industrial (non-proprietary) 
  server: 
  performance overhead and binary size increase (in $\times$) of  \approach{} 
  compared~to~\CC{} %
  }
  \begin{tabular}{l|ccccc|c|c|c}
    \toprule
    \multirow{2}{*}{\makecell{\textbf{\approach{}}\\ \textbf{Pass}}} & \multicolumn{5}{c|}{\textbf{\A{} (Block Size)}} & \multirow{2}{*}{\textbf{\M{}}} & \multirow{2}{*}{\textbf{\MD{}}} & \multirow{2}{*}{\textbf{\B{}}} \\
    \cmidrule(lr){2-6}
    & 
    (1) & (2) & (4) & (8) & (dep) & & & \\
    \midrule
    \textbf{Performance} & 2.18 & 1.79 & 1.64 & 1.52 & 1.75 & 1.17 & 53.67 & 1.17 \\
    \midrule
    \textbf{Binary Size} & 8.47 & 5.88 & 4.59 & 4.11 & 6.11 & 3.06 & 13.12 & 1.24 \\
    \specialrule{\heavyrulewidth}{0pt}{2pt}
    \multicolumn{9}{l}{\footnotesize\textit{%
    Server: Intel Xeon Gold 6226R, 2.9GHz, 2$\times$16 cores, 2 threads/core,}} \\
    \multicolumn{9}{l}{\footnotesize\textit{hyperthreading on, L1: 1MB D+I, L2: 32MB, L3: 44MB, Mem: 500\,GB}} \\
  \end{tabular}
  \label{tab:overheads-artifact}
\end{table}

The central claim reproduced by this artifact is the per-pass
\textbf{runtime performance overhead} and \textbf{binary size increase} of the
ITHICA transformations relative to the uninstrumented \CC{} baseline,
i.e., the results of Table~\ref{tab:overheads} and Obs.~8 in the paper.
The first goal of the evaluation is to successfully use the ITHICA
passes to compile the two programs; the measurement experiments then
run the resulting binaries. Expected values on the reference host are
given in Tables~\ref{tab:overheads-artifact} and~\ref{tab:fb-overhead};
absolute values vary with the underlying hardware
(\S\ref{sec:artifact-hw}). %
 
\emph{\textbf{Getting started:}} \code{cpu-check-ae/scripts/build\_pass.sh}~builds the  ITHICA pass, e.g., \code{arith|mem|memdiv|br|arithmem} etc.: %
{\aecodesetup\begin{verbatim}
cd cpu-check-ae
scripts/build_pass.sh arith 
\end{verbatim}
}
 
\emph{\textbf{Building the two programs.}} \code{build.sh} compiles
each program with the chosen ITHICA pass. For cpu-check:
{\aecodesetup\begin{verbatim}
cd cpu-check-ae
scripts/build.sh --pass=CC      # baseline, no ITHICA
scripts/build.sh --pass=arith   # ITHICA pass 
# knobs: --blocksize=1|2|4|8|0 --interleaving=1|2|4|8|0
# 0 used for "dep" (blocksize) and "max" (interleaving) 
\end{verbatim}
}
For Fleetbench:
{\aecodesetup\begin{verbatim}
cd fleetbench-ae
scripts/build.sh --bench=hashing --nopass      # baseline
scripts/build.sh --bench=hashing --pass=arith  # ITHICA
\end{verbatim}
}

Since \texttt{proto} and \texttt{stl} take hours to build, pre-built binaries in \code{Binaries/} can be used %
for the next step (``Measuring'') by supplying  \code{-{}-fb-dir=Binaries} to the measuring script.

Each \code{build.sh} \code{-{}-pass=P} (for cpu-check) copies its binary into \code{built/P\_bs<BS>\_intr<I>\_cpu\_check}; each \code{-{}-bench=B} (for \texttt{FB}) run extracts a self-contained bundle (with the benchmark's required runfiles) into \code{built/B\_baseline} or \code{built/B\_arith}.

\emph{Note:} the \code{build.sh} scripts are independent of
\code{build\_pass.sh}: each rebuilds the ITHICA pass itself before
compiling the program, and emits verbose logs, including a ``Running on
function'' message per compiled function   confirming that the pass is
applied throughout the program.
 
\emph{\textbf{Measuring.}} \code{measure.sh} runs the binaries in
the directory specified with \code{-{}-cc-dir=<dir>} (cpu-check) or
\code{-{}-fb-dir=<dir>} (Fleetbench). The directory must contain the
uninstrumented baseline and at least one ITHICA-instrumented binary. 
Pre-built binaries in \code{Binaries/} can be used by supplying \code{-{}-cc-dir=Binaries}~(\S\ref{sec:artifact-notes}) to directly reproduce Table~\ref{tab:overheads-artifact}.  %
 
\emph{\textbf{E1---For cpu-check:}}  performance overhead and
binary size increase (main result; ${\sim}40$\,min for all Table~\ref{tab:overheads}
configurations). %
{\aecodesetup\begin{verbatim}
cd cpu-check-ae
scripts/measure.sh --sec=120 --cc-dir=<built|Binaries>   
# -> results/cpucheck_results.txt   (table)
#    results/logs/*.out             (raw binary outputs)
\end{verbatim}
}
Each binary runs for a fixed time (120 sec). Performance overhead is calculated as the ratio of \CC{}'s to the instrumented binary's throughput of
rounds (\code{``SUCCESSES''} in logs) completed within the specified time.

\emph{\textbf{Expected Result}: } Similar to
Table~\ref{tab:overheads-artifact}, also on par with the average (Avg) ratios
in Table~\ref{tab:overheads}. 
The exact ratios will vary based on the hardware used~(\S\ref{sec:artifact-notes}).  
Binary sizes
might differ too, due to %
logging and structural changes for better user experience, code readability, and homogeneity across passes. 
Binary sizes changed the most for \M{}, \MD{}, and \B{} in the artifact (updated to 4.19$\times$, 5.89$\times$, and 2.25$\times$ respectively). These code changes do not affect performance or the functionality of the passes. We also include on GitHub the original versions of these passes that match the  binary sizes of Tables~\ref{tab:overheads}/\ref{tab:overheads-artifact}, under \code{ithica-artifact/cpu-check-ae/Original\_*}.

\emph{\textbf{E2---For Fleetbench:}} 
same metrics as E1, with vs.\
without \A{} (${\sim}25$\,min for all six benchmarks). %
{\aecodesetup\begin{verbatim}
cd fleetbench-ae
scripts/measure.sh --sec=120 --fb-dir=<built|Binaries>
# -> results/fleetbench_results.txt (perf)
#    results/fleetbench_sizes.txt   (size)
#    results/logs/*.out             (raw binary outputs)
\end{verbatim}
}
Each benchmark reports time per operation; the overhead is the ratio of the instrumented to the baseline median ns/op. 

\emph{\textbf{Expected Result:}} Similar to
Table~\ref{tab:fb-overhead}. Only one sub-benchmark from each of the six benchmark families is reported.

{\renewcommand{\arraystretch}{0.9}
\begin{table}[t]
\footnotesize
\caption{Performance overhead and binary size increase
(in $\times$) of \texttt{FB-Arith} compared to the uninstrumented \texttt{FB} baseline %
Measured on the reference host of Table~\ref{tab:overheads-artifact}. 
}
\centering
\begin{tabular}{lccc}
\toprule
\textbf{Benchmark} & \textbf{Sub-benchmark} & \textbf{Performance}  & \textbf{Binary Size}  \\
\midrule
\texttt{hashing}  & ComputeCrc32c (hot)   & 1.42 & 1.11 \\
\texttt{proto}    & Arena                 & 1.44 & 2.87 \\
\texttt{swissmap} & InsertHit (cold)       & 2.99 & 2.31 \\
\texttt{stl}      & Cord                  & 2.79 & 5.81 \\
\texttt{libc}     & Memcpy                & 2.21 & 1.09 \\
\texttt{tcmalloc} & empirical\_driver     & 1.82 & 1.18 \\
\bottomrule
\end{tabular}
\label{tab:fb-overhead}
\end{table}}

\subsection{Experiment customization}

\begin{itemize}
  \item \emph{Pass selection:} \code{arith}, \code{mem},
        \code{memdiv}, \code{br}, plus the combinations
        \code{arithmem}, \code{arithmemdiv},
        \code{arithmemdivbr}.
  \item \emph{Pass configuration parameters}:
        \code{-{}-blocksize=X}, \\ \code{-{}-interleaving=Y}, with \texttt{X}, \texttt{Y} $\in \{0,1,2,4,8\}$.
  \item \emph{Run duration:} \code{scripts/measure.sh -{}-sec=N}
        sets the per-binary wall-clock budget (default 120). %
  \item \emph{CPU set:} \code{measure.sh} %
  pins one worker per available CPU (\S\ref{sec:artifact-notes}). To do it manually, run \texttt{cpu\_check} with
        \code{-c<list>}.
\end{itemize}

\subsection{Notes}\label{sec:artifact-notes}

\begin{itemize}
\item \emph{Measurement precision and stability.} The exact ratios in Tables~\ref{tab:overheads-artifact} and~\ref{tab:fb-overhead} depend on the underlying hardware~(e.g.,~its core count, hyperthreading and AVX support). %
A~host with at least 32 cores is recommended for comparable results~(\S\ref{sec:artifact-hw}). Ratios can also vary run-to-run on the same hardware, due to frequency scaling or co-location with other~load. %
  
  \item \emph{Thread placement.} cpu-check sizes its worker pool from \code{std::thread::hardware\_concurrency()} and spawns one worker per logical CPU, independent of cgroup/affinity limits. On a restricted host this over-spawns and distorts the measurement. The scripts pass an explicit \texttt{-c} list so exactly one worker runs per available CPU.

  \item \emph{Pre-built binaries.} %
  All evaluated programs come pre-built in  \code{Binaries/}, avoiding the need to rebuild from source. %
  
\end{itemize}

\end{document}